\documentclass{aa}
\usepackage{psfig}
\def\la{\;
\raise0.3ex\hbox{$<$\kern-0.75em\raise-1.1ex\hbox{$\sim$}}\; }
\def\ga{\;
\raise0.3ex\hbox{$>$\kern-0.75em\raise-1.1ex\hbox{$\sim$}}\; }
\newcommand{\zabs}{$z_{\rm abs}\,$}
\newcommand{\zem}{$z_{\rm em}\,$}

\newcommand{\kms}{km~s$^{-1}\,$}
\newcommand{\cm}{cm$^{-2}\,$}
\newcommand{\cmm}{cm$^{-3}\,$}
\begin{document}
\title{Photoionized \ion{O}{vi} absorbers 
toward the bright QSO HE~0515--4414\thanks{Based on observations 
with the NASA/ESA 
Hubble Space Telescope, obtained at the Space Telescope Science Institute,
which is operated by Aura, Inc., under NASA contract NAS 5-2655; and on 
observations collected at
the VLT/Kueyen telescope ESO 
(Paranal, Chile), under programme ID 066.A-0212.
}
}
\author{
S. A. Levshakov\inst{1}
\and I. I. Agafonova\inst{1}
\and D. Reimers\inst{2}
\and R. Baade\inst{2}
}
\offprints{S. A. Levshakov}
\institute{
Department of Theoretical Astrophysics,
Ioffe Physico-Technical Institute,
194021 St. Petersburg, Russia
\and
Hamburger Sternwarte, Universit\"at Hamburg,
Gojenbergsweg 112, 21029 Hamburg, Germany
}
\date{Received 00  / Accepted 00 }
\abstract{We report on detailed Monte Carlo inversion analysis of five
\ion{O}{vi} systems from the spectrum of the bright quasar HE 0515--4414
(\zem = 1.71).
The associated system at \zabs = 1.697 
with the neutral hydrogen column density
$N$(\ion{H}{i}) $= 4.4\times10^{13}$ \cm\,
shows pronounced absorption from
highly ionized transitions of \ion{C}{iii}, \ion{C}{iv}, 
\ion{N}{v}, \ion{O}{vi}, \ion{Si}{iv}, and probably \ion{S}{vi}.
We found that only a power law type ionizing spectrum 
($J_\nu \propto \nu^{-1.5}$)
is consistent with the observed sample of the line profiles, 
i.e. the system is definitely intrinsic.  
The relative metal abundances 
give almost the solar pattern and the metallicity of 
$\sim5$ times solar. 
The system originates in a thin shell of the
line-of-sight 
thickness $L \la 16$ pc.
Two \ion{O}{vi} systems at \zabs = 1.674 ([C/H] $\simeq -1.6$)
and 1.602 ([O/H] $\simeq -1.1$), arising in intervening halos,
have linear sizes of $L \simeq$ 3--14 kpc and $\simeq 17$ kpc, respectively.
Absorption systems at \zabs = 1.385 ([C/H] $\simeq -0.3$, 
$L \simeq$ 1.7--2.5 kpc) 
and \zabs = 1.667 ([C/H] $ \simeq -0.5$, $L \simeq 1$ kpc)
exhibit characteristics very
similar to that observed in metal-enriched high velocity clouds 
in the Milky Way. 
These systems are probably embedded in extremely metal-poor
halos with 
[C/H] $< -2.4$ (\zabs = 1.667) and [C/H] $< -3.7$ (\zabs = 1.385).
We also found two additional 
extremely metal-poor Ly$\alpha$ systems at \zabs = 1.500 and 1.681 with,
respectively,
$N$(\ion{H}{i}) $\simeq 1.7\times10^{15}$ and $1.8\times10^{15}$ \cm\, and 
[C/H] $< -4.0$ and $<-3.0$, --- an indication that the distribution of
metals in the metagalactic medium is utterly patchy.
Our results show that 
the ionization states in the analyzed \ion{O}{vi} absorbers,
ranging from $z \simeq 1.4$ to 1.7,
can be maintained by photoionization only and that
the fraction of the shock-heated
hot gas with temperature $T_{\rm kin} > 10^5$ K is negligible in these systems.   
}
\authorrunning{Levshakov et al.}
\titlerunning{\ion{O}{vi} absorbers toward HE 0515--4414}
\maketitle

\section{Introduction}
 

In recently published paper, Reimers et al. (2001, hereafter RBHL)
reported on the identification of six \ion{O}{vi} absorption systems
in the redshift range $1.38 < z < 1.73$ found in the light of the
luminous quasar HE 0515--4414, $z_{\rm em} = 1.71$, $V = 14.9$
(Reimers et al. 1998).
Here we present the physical parameters and metal abundances
for four of these systems and for a fifth new one.
The analysis was performed
with the Monte Carlo inversion (MCI) algorithm which is described
in detail in our previous publications (Levshakov et al. 2000,
hereafter LAK; and  Levshakov et al. 2002).
Since this technique is relatively new, we  briefly
outline its basics below.

The main assumption of the MCI procedure
is that all lines observed in an intervening absorption system
arise in a {\it continuous} absorbing gas slab of a thickness $L$ with
a fluctuating gas density and a random velocity field.
The metal abundances within the absorber are
constant (the validity of this assumption is discussed in
Levshakov et al. 2003a), 
the gas is in the thermal and ionization equilibrium,
and it is optically thin in the ionizing background radiation.
The intensity and the shape of the photoionizing radiation
are considered as external parameters.

Within the absorbing region the radial velocity $v(x)$ and the total
volumetric gas density $n_{\rm H}(x)$ along the line of sight are
considered as two continuous random functions represented
by their sampled values at equally spaced intervals $\Delta x$.
The computational procedure
is based on the adaptive simulated annealing.
Fractional ionizations of different elements are calculated
at every space coordinate $x$ with
the photoionization code CLOUDY (Ferland 1997).

Using the MCI, we estimate the following physical characteristics
of the intervening gas cloud: the mean ionization parameter $U_0$,
the total hydrogen column density $N_{\rm H}$,
the line-of-sight velocity and
density dispersions of the bulk material
($\sigma_{\rm v}$ and $\sigma_{\rm y}$, respectively),
and the chemical abundances $Z_{\rm a}$ of all elements
involved in the analysis.
With these parameters we can further calculate
the mean gas number density $n_0$,
the column densities for different species $N_{\rm a}$, 
the mean kinetic temperature
$T_{\rm kin}$, and the linear size $L$.
This comprehensive information makes it possible to classify the absorber
more reliably and hence to obtain important clues concerning the
physical conditions in the high redshift objects.

While interpreting results of the MCI,
the following should be taken into account.
For every point within the line profile the observed intensity results
from a {\it superposition} of different ionization states (for details, see 
Fig.~1 in LAK). The value of the mean ionization parameter
$U_0$ is related to the parameters of the gas cloud as 
(see Eq.[28] in LAK)
\begin{equation}
U_0 = \frac{n_{\rm ph}}{n_0}(1+\sigma^2_{\rm y})\; .
\label{eq:E1}
\end{equation}
Here $n_{\rm ph}$ is the number density of photons with energies above 1 Ryd
which is determined by
\begin{equation}
n_{\rm ph} = \frac{4\pi}{c h}J_{912}\int^{\infty}_{\nu_{\rm c}}\,
\left(\frac{J_\nu}{J_{912}}\right)\frac{d\nu}{\nu}\;,
\label{eq:E2}
\end{equation}
where $c, h, \nu_{\rm c}$, $J_\nu$, and $J_{912}$ are, respectively,
the speed of light, the Planck constant, the frequency of the
Lyman limit, the mean specific intensity of the radiation background
averaged over all lines of sight, and the mean specific
intensity at the hydrogen Lyman edge (in units of erg cm$^{-2}$ s$^{-1}$
Hz$^{-1}$ sr$^{-1}$).

Eq. (1) shows that if the density field is fluctuating 
($\sigma_{\rm y} > 0$), then with the same mean density $n_0$ and
the same background ionizing spectrum a higher value of
$U_0$ can be realized
without any additional sources of ionization. 
Intermittent regions
of low and high ionization caused by the density fluctuations will occur
in this case along the line of sight leading to a lower value of
the total hydrogen column density $N_{\rm H}$ as compared with a
homogeneous gas slab model ($\sigma_{\rm y} = 0$).
On the other hand, for a given $U_0$
the mean gas density $n_0$ is also higher in fluctuating media. 
Since $L = N_{\rm H}/n_0$, 
the value of the linear size calculated under
the assumption of a fluctuating gas density is
much smaller as compared to that obtained for a model of a constant 
gas density. 

One of the problem 
in the interpretation of the \ion{O}{vi} absorbers
is a high overabundance of oxygen compared to other elements, in particular,
to carbon and especially to silicon 
(e.g. Carswell et al. 2002)
which is obtained from 
the photoionization models assuming a homogeneous gas density. 
It is well known that the metal abundances measured in the absorption
systems depend in a crucial way on the adopted 
spectral shape of the ultraviolet background ($\lambda < 912$ \AA).
A mean metagalactic spectrum of the UV radiation 
as a function of $z$ may be estimated
from  theoretical calculations
(e.g., Haardt \& Madau 1996, hereafter HM; Fardal et al.1998), 
but for every particular system 
one can never exclude that the shape of this spectrum  
is affected by local sources 
(see, e.g., Reimers et al. 1997; Kriss et al. 2001; Smette et al. 2002).
However, if the system contains metal
lines of many ions in different ionization stages and the relative
abundance ratios of the elements 
are known either 
from observations (e.g., in galaxies or stars) or from 
theoretical calculations of stellar nucleosynthetic yields, then  
it is possible
to restore the spectral shape of the ionizing continuum responsible for
the ionization states in a system under study. 
We used this approach in the
present work while analyzing the absorption systems at \zabs = 1.385, 
1.602, 1.667, 1.674, and 1.697. 

The structure of the paper is as follows. The observations are described
in Sect.~2. Physical states of the \ion{O}{vi} absorption systems
are outlined in Sect.~3.
Notes on individual \ion{O}{vi} absorbers are given in Sect.~4.
The obtained results are discussed in Sect.~5, and we draw our conclusions
in Sect.~6.

\section{Observations}

Observations and data reduction are described in detail in RBHL.
Here we list the main characteristics of the data obtained 
which are relevant to the present study.

HE 0515--4414 was observed with the HST/STIS in January and February, 2000
with the medium resolution NUV echelle mode (E230M) and a $0.2\times0.2$
aperture which provides a resolution of $\sim 30000$ (FWHM $\simeq 10$
\kms). The spectrum covers the range between 2279 \AA\, and 3080 \AA\,
where the moderate S/N ratio (between 10 and 20 per resolution 
element) is achieved.

Echelle spectra of HE 0515--4414 were obtained during 
ten nights between October 7, 2000 and January 2, 2001 using
the UV-Visual Echelle Spectrograph (UVES) installed 
at the VLT/Kueyen telescope. These observations were carried
out under good seeing conditions 
(0.47--0.70 arcsec)
and a slit width of 0.8 arcsec giving
the spectral resolution of 
$\sim 55000$
(FWHM $\simeq 6$ \kms). 
The VLT/UVES data have very high S/N ratio 
(between 50 and 100 per resolution
element) which allows us to detect very weak absorption features.

The observed wavelength scales for both sets of data were transformed
into vacuum heliocentric wavelengths. The calibrated spectra
were added together using weights proportional to their S/N.

\begin{figure}
\vspace{0.0cm}
\hspace{0.0cm}\psfig{figure=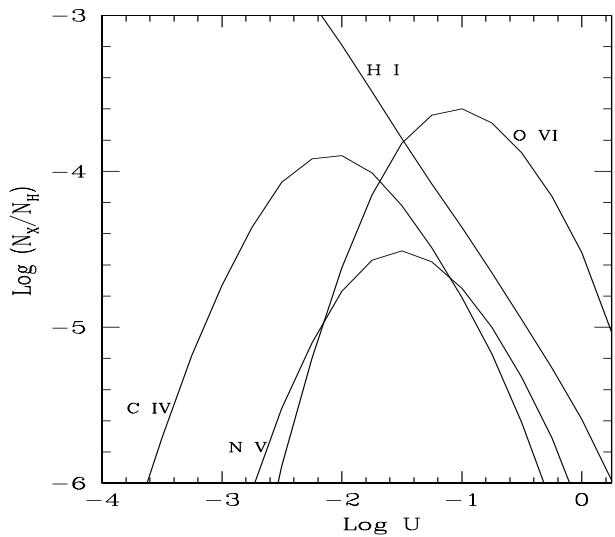,height=8.0cm,width=8.0cm}
\vspace{-0.7cm}
\caption[]{
Relative column densities (in units of the total hydrogen
column density $N_{\rm H}$) plotted as a function of the
ionization parameter $U$ for a highly ionized
gas in photoionization equilibrium 
(the UV background field $\propto \nu^{-1.5}$).
Photospheric solar abundances
from Grevesse \& Sauval (1998) are used.
}
\label{fig1}
\end{figure}

\section{\ion{O}{vi} absorption in photo- and shock-heated gas}

The observation and interpretation of the extragalactic \ion{O}{vi}
absorption systems is closely linked to cosmological simulations
of hierarchical structure formation
which predict that a substantial fraction of intergalactic gas is
shock-heated to high temperatures ($T_{\rm kin} \sim 10^5 - 10^7$ K)
at low redshift and that
a considerable fraction of baryons (up to $\sim 50$\%)
resides in this `warm-hot' gas phase at $z = 0$. 
With increasing redshift, the fraction of baryons 
decreases rapidly (down to $\sim 5$\%)
at $z = 3$ (e.g., Cen \& Ostriker 1999; Dav\'e et al. 2001).
Recently obtained results indicate that 
the cosmological mass density of the
low-$z$ \ion{O}{vi} systems 
may be comparable to the combined cosmological mass density of 
stars and cool gas in galaxies and X-ray gas in galaxy clusters
at low $z$ (Tripp et al. 2000).

The interpretation of the \ion{O}{vi} lines is not, however,
straightforward, especially when we are dealing with high-$z$ systems. 
The resonance line doublet of \ion{O}{vi} may arise 
in both collisionally ($T > 10^5$ K)
and photoionized ($T \sim 10^4$ K) gas. Moreover,
the absorbing region may be not homogeneous, 
a gas slab may be exposed to a time-dependent source of ionizing
radiation, and what we
observe may be a mixture of different phases, which may not
be in ionization equilibrium  (e.g., regions behind shock fronts).

To outline the main characteristics of an absorption system arising in the
collisionally or radiatively ionized gas, 
it is, however, worthwhile to consider an equilibrium situation.
A non-equilibrium case should lie between the results obtained for
collisional or radiative equilibrium.
In this regard, it is very illustrative to
compare column densities of the most abundant ions 
calculated for a simplified model
of a plane parallel gas slab that is either in photoionization equilibrium
in a radiation field 
or in collisional
equilibrium at a given temperature $T$. The gas is assumed to be optically
thin in the ionizing Lyman continuum.

\begin{figure}
\vspace{0.0cm}
\hspace{0.0cm}\psfig{figure=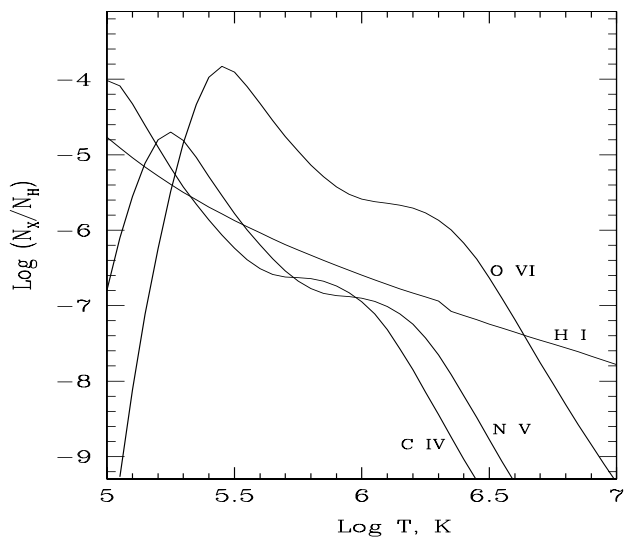,height=8.0cm,width=8.0cm}
\vspace{-0.7cm}
\caption[]{
Relative column densities (in units of the total hydrogen
column density $N_{\rm H}$) plotted as a function of the
temperature $T$ for a highly ionized
gas in collisional ionization equilibrium. Photospheric solar abundances
from Grevesse \& Sauval (1998) are used.
}
\label{fig2}
\end{figure}

If we define $N_{{\rm a},i}$  as
the total ion column density of element `a' in the $i$th ionization
stage ($N_{\rm a} = \sum_i\,N_{{\rm a},i}$) and 
$ \Upsilon_{{\rm a},i} = N_{{\rm a},i}/N_{\rm a}$
as the fractional ionization of ions $\{ {\rm a},i \}$,
then their relative column density in units of
the total hydrogen column density $N_{\rm H}$ is given by
\begin{equation}
N_{{\rm a},i}/N_{\rm H} = Z_{\rm a}\,\Upsilon_{{\rm a},i}\; ,
\label{eq:E3}
\end{equation}
where $Z_{\rm a}$ is the metal abundance.

Figs.~1 and 2 show the relative column densities
of \ion{H}{i}, \ion{C}{iv}, \ion{N}{v}, and \ion{O}{vi} 
(for illustration,
solar abundances are used)
plotted against
the ionization parameter $U$ for the photoionization caused by the
power law background $J_\nu \propto \nu^{-1.5}$ 
(calculated with CLOUDY), and against
$T$ for the collisional ionization case (the ionization fractions 
are
from Sutherland \& Dopita 1993).
In both cases \ion{O}{vi} dominates for high excitation conditions.
The photoionization equilibrium is characterized by 
strong hydrogen lines in the range $\log U \la -1$ where all metal
absorptions can be observed simultaneously.
In collisional ionization equilibrium Ly$\alpha$ is much weaker 
in the range $5.5 \la \log (T) \la 6.0$ where  
\ion{O}{vi} is the most readily observed ion in the ultraviolet range.
At these temperatures, other
lines (\ion{C}{iv} and/or \ion{N}{v}) are also very weak 
and hardly observable in real spectra.
Indeed, if we assume an absorber of $N_{\rm H} = 10^{18}$ \cm,
then at $T_{\rm kin} = 3\times10^5$ K the central optical depths of these
species and their Doppler $b$-parameters are as follows:
$\tau_0$(Ly$\alpha$) = 0.014, $b_{{\rm Ly}\alpha} = 73$ \kms;
$\tau_0$(\ion{C}{iv}$_{1548}$) = 0.012, $b_{{\rm C IV}}$ = 21 \kms;
$\tau_0$(\ion{N}{v}$_{1238}$) = 0.015, $b_{{\rm N V}}$ = 20 \kms; and
$\tau_0$(\ion{O}{vi}$_{1031}$) = 1.40, $b_{{\rm O VI}}$ = 18 \kms.

For the abundances different from the solar values the estimated
optical depths of the metal lines can be scaled correspondingly using
Eq.(\ref{eq:E3}) and the curves from Fig.~2.

\begin{figure}
\vspace{0.0cm}
\hspace{0.0cm}\psfig{figure=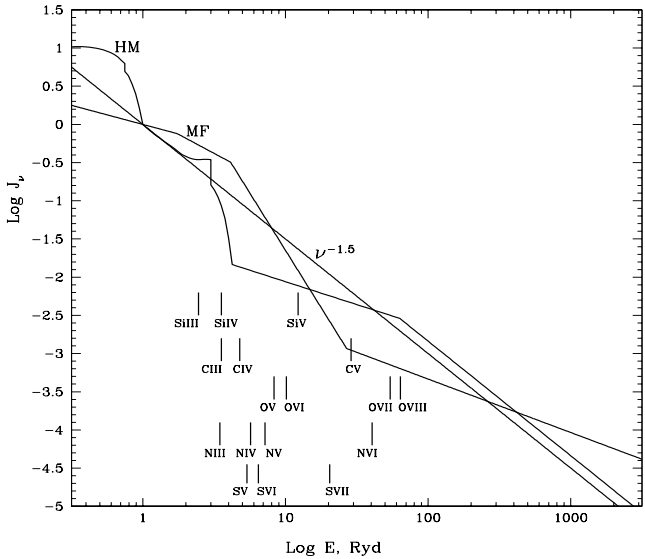,height=8.0cm,width=8.0cm}
\vspace{-0.5cm}
\caption[]{UV-background ionizing continua used in the present
calculations. The spectra have been normalized
so that $J_\nu$($h\nu=1$ Ryd) = 1.0. 
Shown are: HM --- a mean metagalactic
spectrum at $z = 2$ with a decrement at 4 Ryd to account for
\ion{He}{ii} attenuation in intervening Ly$\alpha$ clouds as 
computed by Haardt \& Madau (1996);
MF --- an AGN type spectrum deduced by
Mathews \& Ferland (1987), and $\nu^{-1.5}$ ---
a single power law spectrum $J_\nu \propto \nu^{-1.5}$.
The positions of ionization thresholds of different ions
are indicated by tick marks.
}
\label{fig3}
\end{figure}

\section{Results on individual \ion{O}{vi} systems}

In this section we consider \ion{O}{vi} systems observed in the
spectrum of HE 0515--4414.
All physical parameters listed below in Table~1 were derived using the
MCI procedure as described in LAK.
The internal errors of the fitting parameters $U_0, N_{\rm H},
\sigma_{\rm v}, \sigma_{\rm y}$ and $Z_{\rm a}$ are about 15\%--20\%,
the errors of the measured column densities are less than 5\% for
the high S/N UVES data and of about 15\%--25\% for
the moderate S/N STIS spectra. 
It should be noted that the errors of the fitting parameters 
reflect merely the configuration
of the parameter space in the vicinity of a minimum of the objective
function (this function is defined by Eqs.[29] and [30] in LAK). 
To what extent the recovered
parameters may correspond to their real values is discussed separately
for each individual absorption system.

In some systems with rather high neutral hydrogen column
densities [$N$(\ion{H}{i}) $\sim 10^{15}$ \cm] 
we did not detect any transitions of high or low most abundant
ions in the observational range from 2279 \AA\, to 6662 \AA.
We will call such systems formally as `metal-free' throughout the text
with all reserve for a probability that it may contain transitions
in the far UV-range like 
\ion{O}{iv} $\lambda 554.1$, \ion{O}{v} $\lambda 629.7$, \ion{O}{vii} $\lambda 21.6$,
\ion{O}{viii} $\lambda 19.0$ or \ion{Ne}{vii} $\lambda465.2$, and \ion{Ne}{viii}
$\lambda770.4$ \AA.
The upper limits on the column densities of ions from the observational range 
were calculated in the following way. 
We define the equivalent width
detection limit, $\sigma_{\rm lim}$, as (Levshakov et al. 1992):
\begin{equation}
\sigma_{\rm lim} = \Delta \lambda \left(\langle {\rm S/N} \rangle^{-2}
{\cal M} + \delta^2_c {\cal M}^2\right)^{1/2}\; .
\label{eq:E4}
\end{equation}
Here $\langle {\rm S/N} \rangle$ is the mean signal-to-noise ratio at the
expected position of an absorption line, $\delta_c = \sigma_c/C$ is the
accuracy of the local continuum fit over the width of the line,
${\cal M} = 2.5\times{\rm FWHM}$ is the full width (in pixels) at the
continuum level of a weak absorption line, and $\Delta \lambda$ is
the pixel size in \AA. In calculations we used 
${\cal M} = 10$, i.e. $\Delta \lambda/\lambda = \frac{1}{4}$FWHM/c.

Eq.(\ref{eq:E4}) shows that
the minimum detectable equivalent width, defined as $W_{\rm lim} =
3\,\sigma_{\rm lim}$, is very sensitive to the accuracy of the
continuum fitting, and $W_{\rm lim}$ increases as the data become more noisy.

For a weak absorption line ($\tau_0 \ll 1$)
the column density does not depend on the broadening parameter $b$ and is
proportional to $W_{\rm lim}$:
\begin{equation}
N_{\rm lim} = 1.13\times10^{20}\,\frac{W'_{\rm lim}}{\lambda^2_0\,f}\, {\rm
cm}^{-2}\; ,
\label{eq:E5}
\end{equation}
where $f$ is the oscillator strength for the absorption transition,  
$\lambda_0$ is the line-center rest wavelength in \AA,
and $W'_{\rm lim}$ is the rest-frame equivalent width. 
Using this linear approximation to the curve of growth, we calculated the
upper limits $N_{\rm lim}$ listed in Table~2 
(absorption lines used in this analysis are listed in Col. [2]).
The values of $\lambda_0$ and 
$f$ were taken from Morton (1991) and from Spitzer \& Fitzpatrick (1993) for
the \ion{Si}{ii} $\lambda1526.7$ \AA\, line.

We also note that the
quasar HE 0515--4414, being the brightest known QSO at
$z > 1.5$, has the
absolute magnitude $M \simeq -30$ (assuming $\Omega = 1$ and $H_0 = 75$
\kms~Mpc$^{-1}$) and, thus, its luminosity  is ${\cal L}_{\rm Q} \simeq
2.5\times10^{47}$ erg s$^{-1}$. 
Three \ion{O}{vi} systems --- that at \zabs = 1.667, 1.674 and 1.697 ---
are not far from the QSO emitting region ($\Delta v_{\rm em-abs} \simeq
4760, 4000$ and 1500 \kms, respectively), and hence 
the intensity of the incident continuum emitted by the
quasar may exceed the value of a mean intensity of the background
ionizing radiation at the positions of these absorbers.  
According to HM, the Lyman-limit specific intensity
$J_{912}$ of the metagalactic
field at $z \simeq 1.6 - 1.8$
is about $0.5\times10^{-21}$ erg cm$^{-2}$ s$^{-1}$ Hz$^{-1}$ sr$^{-1}$.
If the velocity differences
$\Delta v_{\rm em-abs}$ are caused entirely 
by the cosmological expansion, then
the proper luminosity distances between the QSO and the absorbers
at \zabs = 1.667, 1.674 and 1.697 are 
14.6\,$h^{-1}_{75}$ Mpc,
12.2\,$h^{-1}_{75}$ Mpc
and 4.3\,$h^{-1}_{75}$ Mpc, respectively, and the corresponding
local Lyman-limit specific intensities due to the QSO at $z = 1.67$ is
$J_{\rm Q} \simeq 2.0\times10^{-21}$ 
erg cm$^{-2}$ s$^{-1}$ Hz$^{-1}$ sr$^{-1}$\,
and $J_{\rm Q} \simeq 1.0\times10^{-20}$
erg cm$^{-2}$ s$^{-1}$ Hz$^{-1}$ sr$^{-1}$ 
at $z = 1.70$
(assuming a single power law spectrum for the
quasar continuum at $\lambda < 912$ \AA\, with the spectral
index $\alpha = -1.5$).
The value of $J_{\rm Q}$ becomes comparable with 
$J_{912}$ only for the \zabs = 1.602 system where 
$J_{\rm Q} \simeq 0.3\times10^{-21}$ 
erg cm$^{-2}$ s$^{-1}$ Hz$^{-1}$ sr$^{-1}$. 
These estimated  $J_{\rm Q}$ values were used to calculate 
the mean densities $n_0$ and linear sizes $L$ of the \ion{O}{vi} 
systems. 

\begin{table*}
\centering
\caption{
Physical parameters of the \ion{O}{vi} systems toward HE 0515--4414
($z_{\rm em} = 1.71$) derived by the MCI procedure
}
\label{tbl-1}
\begin{tabular}{lccccccc}
\hline
\noalign{\smallskip}
Parameter &\multicolumn{2}{c}{\zabs = 1.385} & \zabs = 1.602 & \zabs = 1.667
& \multicolumn{2}{c}{\zabs = 1.674} & \zabs = 1.697\\
(1) & (2) & (3) & (4) & (5) & (6) & (7) & (8)\\
\noalign{\smallskip}
\hline
\noalign{\smallskip}
UV$^a$ & mHM & HM & HM & $\nu^{-1.5}$ & HM & $\nu^{-1.5}$ & $\nu^{-1.5}$\\
$U_0$ & 0.11& 0.11 & 0.26 & 0.20 & 0.18 & 0.07 & 0.06\\
$N_{\rm H}$, cm$^{-2}$ & 1.2E18 & 9.9E17 & 8.1E18 & 1.1E18  
& 5.5E19 & 1.7E19 & 4.8E17 \\
$\sigma_{\rm v}$, km~s$^{-1}$ & 18.8 & 18.8 
& 29.3 & 14.0 & 53.7 & 53.0& 19.2 \\
$\sigma_{\rm y}$ & 0.4 & 0.5 & 0.9 & 0.3 & 0.9 & 0.9 & 0.4\\
\noalign{\smallskip}
$Z_{\rm C}^b$ &1.2E-4 & 9.3E-5 & $<$7.5E-6 & 7.0E-5 &
1.8E-6 & 5.7E-6 & 1.3E-3\\
$Z_{\rm N}$ &$<$2.1E-5 & $<$2.1E-5 & $<$2.7E-5 & 1.2E-5 &
$<$3.0E-7 & $<$5.0E-7 
& 3.0E-4 \\
$Z_{\rm O}$ &3.8E-4 & 4.5E-4 & 4.4E-5 & 1.7E-4 &
1.9E-5 & 4.1E-5 & 3.3E-3 \\
$Z_{\rm Si}$&$<$3.5E-5 & $<$3.5E-5 & $\ldots$ & $\ldots$ &
$<$1.3E-6 & $<$2.0E-6 & 2.3E-4 \\
$Z_{\rm S}$ & $\ldots$ & $\ldots$ & $\ldots$ &
$\ldots$ & $\ldots$ & $\ldots$ & 8.0E-5\\
$[Z_{\rm C}]^c$&$-0.31\pm0.09$&$-0.42\pm0.08$&$<-1.51$& $-0.54\pm0.15$ &
$-2.13\pm0.09$ & $-1.63\pm0.09$ & $0.72\pm0.08$ \\
$[Z_{\rm N}]$&$<-0.61$ & $<-0.61$ & $<-1.51$ & $-0.85\pm0.15$ 
& $<-2.41$ & $<-2.21$ &$0.55\pm0.12$\\
$[Z_{\rm O}]$&$-0.11\pm0.17$&$-0.04\pm0.17$&$-1.05\pm0.23$& $-0.46\pm0.20$ 
& $-1.41\pm0.12$ & $-1.08\pm0.12$ & $0.83\pm0.11$ \\
$[Z_{\rm Si}]$&$<0$ & $<0$ & $\ldots$ & $\ldots$ &
$<-1.4$ & $<-1.2$ & $0.84\pm0.15$\\
$[Z_{\rm S}]$&$\ldots$&$\ldots$&$\ldots$ & $\ldots$ & $\ldots$ & $\ldots$
& $0.57\pm0.15$ \\
\noalign{\smallskip}
$N$(H\,{\sc i}), \cm&5.6E13 & 5.3E13 & 1.4E14 & 2.2E13 & 
9.3E14 & 1.3E15 & 4.4E13 \\
$N$(C\,{\sc iii}), \cm&7.0E12 & 5.5E12 & $<$1.3E12 & $\ldots$ & 
$\ldots$ & $\ldots$ & 9.0E12 \\
$N$(N\,{\sc iii}), \cm&$\ldots$&$\ldots$&$\ldots$&$\ldots$&
$\ldots$&$\ldots$ & $<$2.1E12\\
$N$(Si\,{\sc iii}), \cm&$\ldots$&$\ldots$&$\ldots$&$\ldots$&
$\ldots$&$\ldots$ & $<$4.0E10\\
$N$(C\,{\sc iv}), \cm&1.8E13 & 1.7E13 & $<$3.5E12&1.4E12& 
9.1E12 & 8.2E12 & 6.8E13 \\
$N$(Si\,{\sc iv}), \cm&$<$1.0E11&$<$1.0E11&$\ldots$&
$\ldots$&$<$6.4E10&$<$7.0E10&2.4E11 \\
$N$(N\,{\sc v}), \cm&$<$4.9E12&$<$4.9E12 & $<$2.9E12&1.4E12 & 
$<1.6$E12 &$<$1.8E12&4.4E13 \\
$N$(O\,{\sc vi}), \cm&1.0E14 & 7.6E13 & 5.1E13 & 5.0E13 & 
1.1E14 & 1.1E14 & 4.5E14 \\
$N$(S\,{\sc vi}), \cm&$\ldots$ & $\ldots$ & $\ldots$ & $\ldots$ & $\ldots$ 
& $\ldots$ & 5.7E12\\
\noalign{\smallskip}
$n_0$, cm$^{-3}$&1.5E-4 & 2.0E-4 & 1.6E-4 & 4.1E-4 & 8.3E-4 & 
2.1E-3 & $\geq$1.0E-2$^d$\\
$\langle T_{\rm kin} \rangle$, K&2.5E4 & 2.3E4 & 4.3E4 & 3.3E4 & 
4.5E4 & 3.7E4 & 1.9E4 \\
$T^{\rm min}_{\rm kin}$, K&1.8E4 & 1.9E4 & 3.0E4 & 2.9E4 & 3.0E4 & 2.6E4 
& 1.6E4 \\
$T^{\rm max}_{\rm kin}$, K&2.7E4 & 2.8E4 & 5.3E4 & 3.8E4 & 7.5E4 & 5.3E4 
& 2.2E4 \\
$L$, kpc&2.5 & 1.7 & 17 & 0.9 & 14 & 2.7 & $\leq$1.6E-2$^d$ \\
\noalign{\smallskip}
\hline
\noalign{\smallskip}
\multicolumn{8}{l}{$^a$HM (mHM) is a Haardt-Madau (modified Haardt-Madau)
type spectrum (see text);}\\
\multicolumn{8}{l}{$^bZ_{\rm X}$ = X/H,\,
$[Z_{\rm X}] = \log (Z_{\rm X}) - \log (Z_{\rm X})_\odot$. }\\
\multicolumn{8}{l}{$^c$Errors of $[Z_{\rm X}]$ include uncertainties in the
solar abundances as well;}\\
\multicolumn{8}{l}{$^d$More probable values of $n_0$ and $L$ are, respectively,
0.2 \cmm\, and 0.8 pc as discussed in Sect.~5}
\end{tabular}
\end{table*}

In our study we used different ionizing background spectra, some of
which are shown in Fig.~3. For every system, we started 
calculations with a HM spectrum, and if it failed to reproduce
self-consistently the observational data  another types of 
photoionizing continua were tested.

\subsection{\ion{O}{vi} absorber at \zabs = 1.697}

The system exhibits a wealth of ions. Namely, at
$N$(\ion{H}{i}) = $4.4\times10^{13}$ \cm\, (see Table~1) we observe
absorption lines of 
\ion{C}{iii}, \ion{C}{iv}, \ion{N}{v},
\ion{O}{vi}, \ion{Si}{iv}, and probable \ion{S}{vi}\,
(as far as we know the ion
\ion{S}{vi} has never been observed in
low \ion{H}{i} column density systems at high redshift).
Besides, there are
continuum `windows' at the positions of the \ion{C}{ii} 1334 \AA,
\ion{Si}{ii} 1260 \AA, 
\ion{N}{iii} 989 \AA,  and \ion{Si}{iii}  1206 \AA\, lines
which can be used to constrain the ionization state in the cloud.
Thus we can expect that the ionization corrections and, hence, 
the metal abundances for all elements
will be estimated with a sufficiently high accuracy. 
This gives us a unique opportunity
to restore the shape of the local ionizing continuum if we further
take into account the measurements 
of the relative abundances in
spiral and elliptical galaxies. For instance,
 the abundance ratios of $\alpha$-elements
(in our case Si and S) to oxygen
appear to be universally constant and independent on metallicity 
(Henry \& Worthey, 1999).

\begin{figure*}
\vspace{0.0cm}
\hspace{0.0cm}\psfig{figure=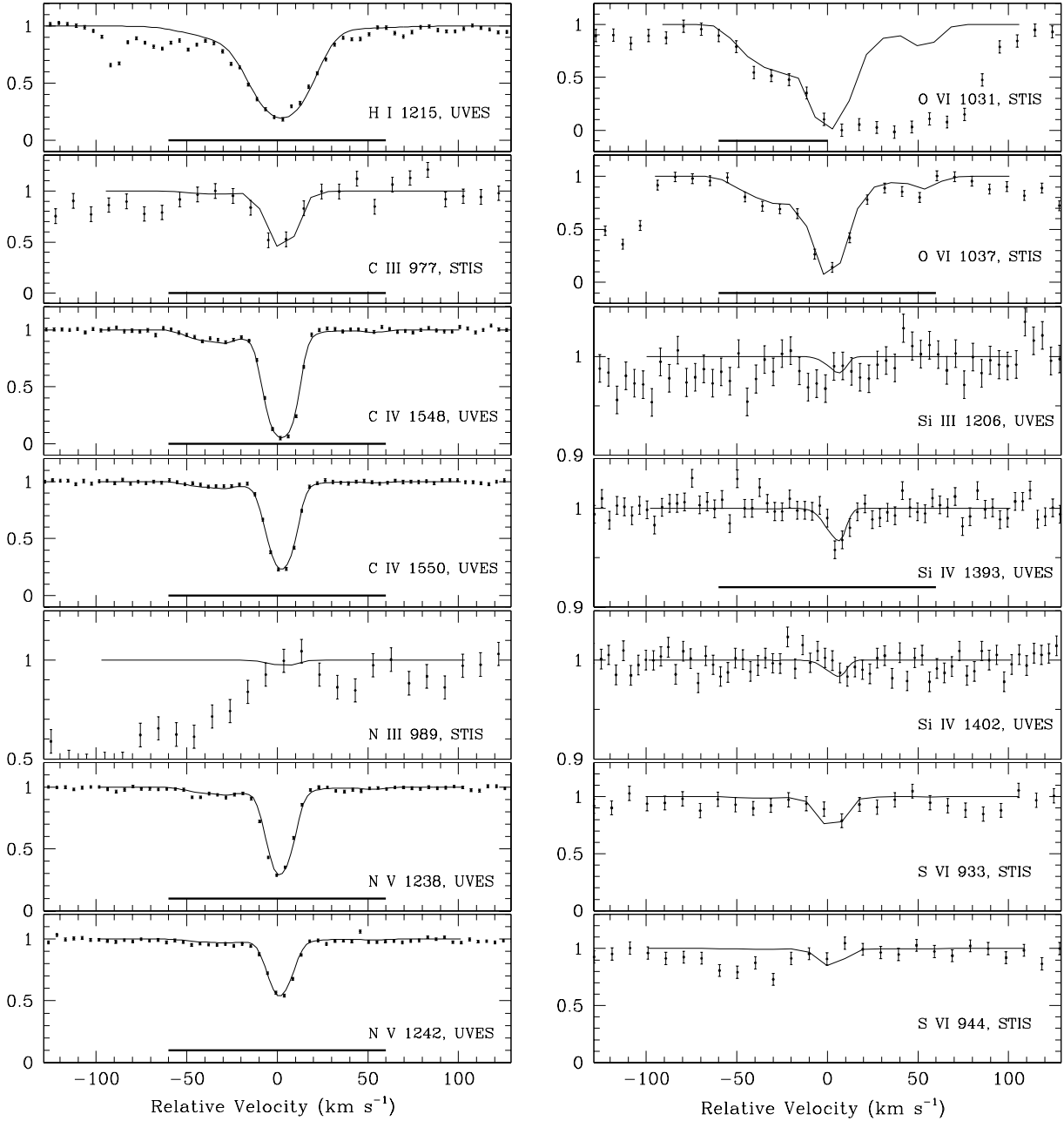,height=18.0cm,width=18.0cm}
\vspace{-1.0cm}
\caption[]{Hydrogen and metal lines associated with the
\zabs = 1.697 \ion{O}{vi} absorption system toward HE 0515--4414
(normalized intensities are shown by dots with $1\sigma$ error bars).
The zero radial velocity is fixed at $z = 1.6971$. Smooth lines are
the synthetic spectra convolved with the corresponding point-spread
functions and computed with the physical
parameters from Table~1, Col. (8). 
Bold horizontal lines mark pixels included
in the optimization procedure. Spectra obtained with the VLT/UVES
and the HST/STIS spectrographs are marked by UVES and STIS, respectively.
The normalized $\chi^2_{\rm min} = 1.10$ (the number of degrees of
freedom $\nu = 276$).
}
\label{fig4}
\end{figure*}

The metal lines in the \zabs = 1.697 cloud 
are completely resolved with both
STIS and UVES spectral resolutions. The line profiles
of different ions at different ionization stages 
demonstrate similar narrow cores, FWHM $\simeq 20$ \kms.  
This fact unambiguously indicates that the turbulent broadening 
dominates over the thermal one,
and as a consequence 
the kinetic temperature of the gas containing highly ionized ions
must be considerably lower than $1.3\times10^5$ K --- the value 
estimated from the width of the \ion{O}{vi} 1037 \AA\, assuming 
a pure thermal line width. 
However, even if $T_{\rm kin} = 1.3\times10^5$ K, then the
ionization fractions of \ion{N}{v} and \ion{O}{vi} in the
collisional ionization equilibrium are  
$\Upsilon_{{\rm N V}} = 0.034$
and $\Upsilon_{{\rm O VI}} = 1.1\times10^{-5}$, respectively 
(e.g., Spitzer 1978; Sutherland \& Dopita 1993). 
The measured column densities $N$(\ion{N}{v}) =
$4.4\times10^{13}$ \cm\, and  $N$(\ion{O}{vi}) = 
$4.5\times10^{14}$ \cm\, would require a unrealistic oversolar
abundance ratio [O/N]\footnote{
[X/Y] $\equiv$ log (X/Y) -- log (X/Y)$_\odot$ .
Throughout the text photospheric solar abundances 
for C and O are taken from Allende Prieto et al. (2001, 2002),
for N and Si from Holweger (2000), 
and for S from Grevesse \& Sauval (1998).
}
= 3.74. 
Thus we can conclude that the fraction of the
collisionally ionized gas is negligible
in the \zabs = 1.697 cloud and that
the ionization state 
within this cloud should be determined by photoionization only.

An attempt to describe this system using a HM
ionizing spectrum at $z = 1.7$ gives inconsistent
profiles of the \ion{C}{iii} 977 \AA\, 
and \ion{C}{iv} 1548, 1550 \AA\, lines: the observed
amount of \ion{C}{iii} is much less than the computed one.
Moreover, the relative abundance ratio [C/O]
was too low: [C/O] $< -0.7$ which is in clear contradiction with 
the observations of extragalactic \ion{H}{ii}  regions
showing [C/O] $\ga -0.5$ 
(Henry et al. 2000). 
Thus, the HM spectrum has been ruled out.

Since this system is probably located not far from the quasar, 
its ionization state can be directly affected by the QSO emission.
Unfortunately, the exact shape of the far UV continuum of  
HE 0515--4414 is not known (it was observed only in the rest frame
region $\lambda > 740$ \AA, see Reimers et al. 1998). 
To approximate the ionizing continuum
we used several power law type
spectra ($\alpha$ ranges from $-1.2$ to $-1.8$) as well as an AGN
type spectrum deduced by Mathews \& Ferland (1987, hereafter MF).
The MF continuum was also rejected because it produced an anomalously
low Si abundance inconsistent with the relative 
(to solar) abundances of other
$\alpha$-elements O and S. 
Among the power law continua the optimal fitting
was found with $\alpha = -1.5$.

\begin{figure}
\vspace{0.0cm}
\hspace{0.0cm}\psfig{figure=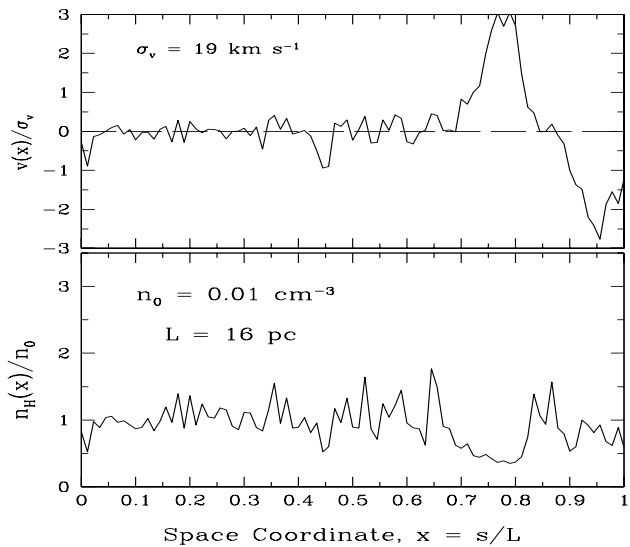,height=8.0cm,width=8.0cm}
\vspace{-0.4cm}
\caption[]{Computed velocity (upper panel) and gas density (lower panel)
distributions along the line of sight within the \zabs = 1.697 absorber
toward HE 05150--4415. Shown are patterns rearranged according to the
principle of minimum entropy production rate (Levshakov et al. 2002).
The $n_0$ and $L$ values shown in the lower panel correspond to
a limiting case which assumes that the difference
$\Delta z_{\rm em-abs}$ is entirely due to cosmological expansion. 
}
\label{fig5}
\end{figure}

The obtained results are listed in Table~1, Col. (8) 
and illustrated in Figs.~4 and 5.
In Fig.~4, parts of line profiles included in the least-squares
minimisation are marked by horizontal lines in the corresponding
panels. The synthetic profiles of the \ion{N}{iii} 989 \AA, 
\ion{Si}{iii} 1206 \AA\, and \ion{S}{vi} 933 \AA\, lines
were calculated in a second round using the 
velocity $v(x)$ and gas density $n_{\rm H}(x)$
distributions already obtained (see Fig.~5) and 
the metallicities chosen in such a way that the synthetic spectra
did not exceed 1 $\sigma$ deviations from the observed normalized
intensities. 

The identification of \ion{S}{vi} should be considered, however,
with some caution because of the limited S/N in the STIS data
which prevents a clear detection of
the weaker 944 \AA\, component of the \ion{S}{vi}
doublet. The observed equivalent width of the stronger 933 \AA\,
component  is $W_{\rm obs} \simeq 52$ m\AA. The STIS data provide the
equivalent width detection limit in this spectral range of
$\sigma_{\rm lim} \simeq 12$ m\AA\, 
if the accuracy of the
local continuum 
fitting $\delta_c$ is
about 5\% and S/N = 10. 
It is seen that formally $W_{\rm obs} > 4\sigma_{\rm lim}$ and
thus the absorption feature at the expected position of the
\ion{S}{vi} 933 \AA\, line is, probably, real. We suggested that
this feature may be attributed to \ion{S}{vi} and found that the
derived sulphur abundance was in line with 
the abundances of other $\alpha$-chain elements. 

Fig.~4 shows that the observed profiles (portions free
from blending) are well represented.
The red wing of the \ion{O}{vi} 1031 \AA\, is partly contaminated
by some Ly$\alpha$ forest absorption, but the \ion{O}{vi}
1037 \AA\, is well fitted to the observed profile. 

\begin{figure*}
\vspace{0.0cm}
\hspace{0.0cm}\psfig{figure=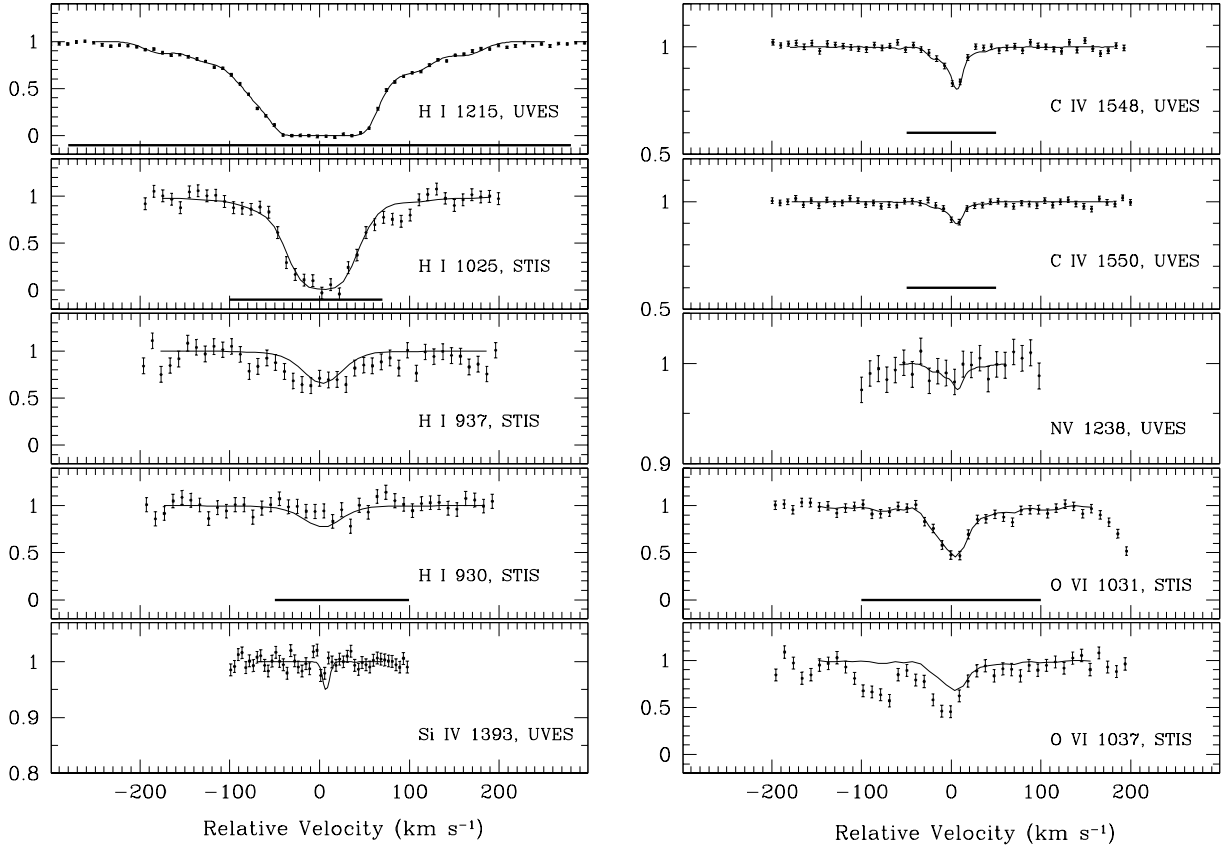,height=18.0cm,width=18.0cm}
\vspace{-6.7cm}
\caption[]{Same as Fig.~4 but for the \zabs = 1.674 \ion{O}{vi} 
absorption system.
The zero radial velocity is fixed at $z = 1.6736$. 
The corresponding physical parameters are listed in Table~1,
Col. (7). 
Here $\chi^2_{\rm min} = 0.93$, $\nu = 487$.
}
\label{fig6}
\end{figure*}

A pronounced asymmetry seen in the wings of the \ion{C}{iv},
\ion{N}{v}, and especially \ion{O}{vi} lines is caused
by the underdense region with the space coordinates $0.65 \leq x \leq 1.0$
where a strong velocity gradient is detected (see Fig.~5).
This figure also shows no high perturbations in the velocity
and density fields along the line of sight
between $x=0$ and $x=0.65$. 

A linear size $L \leq 16$ pc was estimated under the assumption that
the velocity difference between the QSO and the \ion{O}{vi}
system is 
entirely due to the cosmological expansion. If the absorbing cloud has its
own significant radial velocity, it could be situated nearer 
to the emitting region (this will reduce the linear size of the cloud
and increase its volumetric density).
The size of this system is in line with
measurements of other associated systems which show the line of sight
extension of order of magnitude from parsecs  to  tens of parsecs 
(e.g., Rauch et al. 1999; Papovich et al. 2000; 
Hamann et al. 2001).

The relative metal abundances measured in this system 
show almost the solar pattern but the metallicity is 
about 
5 times solar.
The total hydrogen column density could be higher and, hence, the
general metallicity lower if the QSO continuum revealed some strong
emission lines at $\lambda \leq 912$ \AA. But the observed spectrum
of HE 0515--4414 does not show any strong emission in this region
(see Reimers et al. 1998). Another possibility to enhance the
hydrogen column density could be due to incomplete covering of 
the continuum source by the absorbing cloud which makes the saturated
lines look like unsaturated ones (e.g., Petitjean et al. 1994;
Petitjean \& Srianand 1999). Obviously it is not the case for the
system under study: none of the observed profiles show flattened bottoms
and
the doublet ratios for the \ion{C}{iv} 1548, 1550 \AA\, and \ion{N}{v}
1238, 1242 \AA\, pairs are fulfilled perfectly.

Near solar or over solar metallicities are frequently 
measured in QSO environments (Hamann \& Ferland 1999)
and in the associated systems
(e.g., Petitjean et al. 1994; Savaglio et al. 1997; Hamann et al. 2000;
Papovich et al. 2000; Hamann et al. 2001).
High metallicity is usually
explained by rapid and extensive star formation in the galaxy 
surrounding the quasar. 
Extremely high metallicity derived from the \zabs = 1.697 
system suggests its
close physical association with the quasar/host galaxy. This system
is probably a product of the blowout related to a starburst and/or
the onset of quasar activity. In this regard possible 
implications to the origin
of some high velocity clouds (HVCs) will be discussed in Sect.~5.

\begin{figure*}
\vspace{0.0cm}
\hspace{0.0cm}\psfig{figure=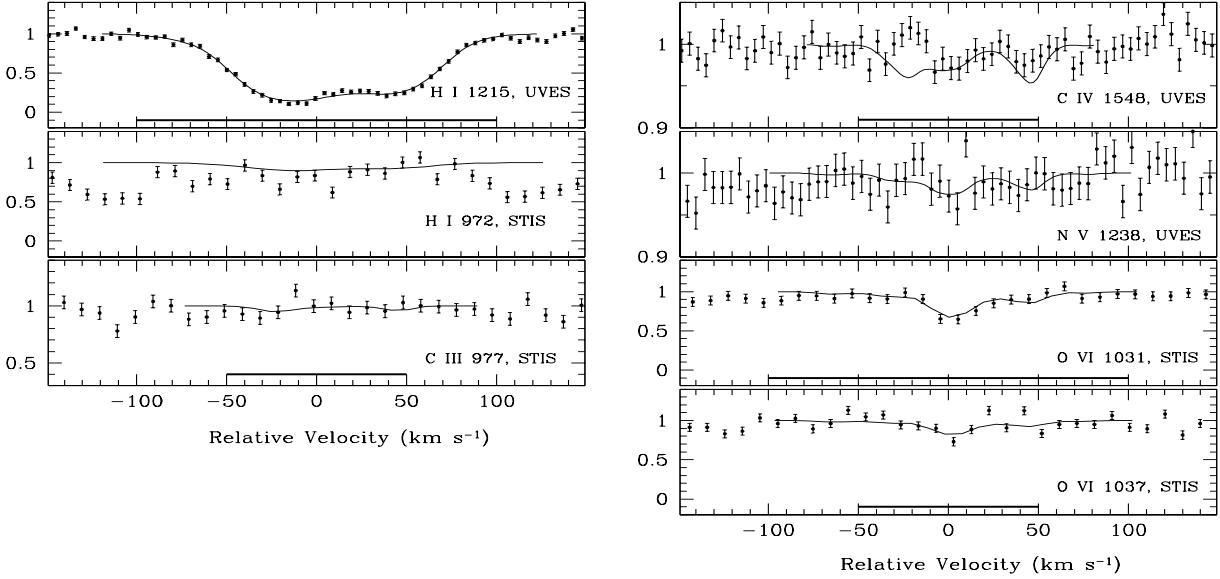,height=15.0cm,width=18.0cm}
\vspace{-7.5cm}
\caption[]{Same as Fig.~4 but for the \zabs = 1.602 \ion{O}{vi} 
absorption system.
The zero radial velocity is fixed at $z = 1.6019$. 
The corresponding physical parameters are listed in Table~1,
Col. (4). 
Here $\chi^2_{\rm min} = 1.00$, $\nu = 163$.
}
\label{fig7}
\end{figure*}

\subsection{\ion{O}{vi} absorber at \zabs = 1.674} 

The absorption system exhibits hydrogen lines up to Ly-7 and 
metal ionic transitions \ion{C}{iv} and 
\ion{O}{vi} (see Fig.~6). 
The Doppler widths of the \ion{C}{iv} and \ion{O}{vi} lines
are $\simeq 20$ \kms\, with $b_{\rm C\,IV}$  slightly lower than
$b_{\rm O\,VI}$ (RBHL). This fact and the co-existence of
essential amounts of \ion{C}{iv} and \ion{O}{vi} requires that
photoionization play a central role in maintaining the
ionization states in this system.
 
At the positions of the \ion{C}{ii} 1334 \AA\, and \ion{Si}{ii} 1260 \AA\,
as well as 
\ion{Si}{iv} 1393 \AA, and \ion{N}{v} 1238 \AA\,
clear continuum windows are seen
which implies that the system is highly ionized (the \ion{C}{iii} 977 \AA\,
line is completely blended).  
The \ion{O}{vi} 1037 \AA\, is blended, but the stronger
1031 \AA\, component is clear and its profile can be very well
fitted to the synthetic spectrum calculated simultaneously
with the carbon \ion{C}{iv} and hydrogen \ion{H}{i} lines.

The Ly$\alpha$ profile has a distinctive shape --- flat bottom of the
width $\simeq 100$ \kms and pronounced 
wings extended over 150 \kms.
It is worth noting that similar spectral shape of Ly$\alpha$
has been observed 
in the \zabs = 4.81 associated system toward the quasar
APM BR J0307--4945 with \zem = 4.753 (Levshakov et al. 2003a). 
Such spectral shapes my be formed in the 
halos of the intervening galaxies with infalling (accretion) 
or outflowing (galactic wind) gas. 

The MCI analysis with the HM spectrum produced results listed in 
Col. (6) of Table~1. The gas density $n_0 = 8.3\times10^{-4}$ \cmm
and the linear size $L = 14$ kpc were calculated with
$J_{\rm Q} = 2.0\times10^{-21}$ 
erg cm$^{-2}$ s$^{-1}$ Hz$^{-1}$ sr$^{-1}$ as estimated above.
This value is 4 times higher as compared 
with $J_{912}$ at $z = 1.7$ given by 
HM, i.e. without additional radiation from the quasar the size of
the absorber would exceed 50 kpc. 

Another type of the UV background was also tested.
Being shifted at only $\simeq 4000$ \kms\, from the  
QSO, the \zabs = 1.674 system may 
be directly influenced by the quasar radiation.
The results obtained
with a power law ionizing continuum ($\alpha = -1.5$,  
$J_{\rm Q} = 2.0\times10^{-21}$ 
erg cm$^{-2}$ s$^{-1}$ Hz$^{-1}$ sr$^{-1}$)
are shown in Fig.~6, whereas
the corresponding physical parameters are given in Col. (7), Table~1. 

The measured abundance pattern [log(C/O), log(N/O), and 
log(O/H)] is similar for both types 
of the UV background spectra
and consistent within the uncertainty range
(the accuracy of our $Z_{\rm C}$ and $Z_{\rm O}$ values 
are about 15\% and 25\%, respectively)
with that observed in the extragalactic \ion{H}{ii}
regions (Henry et al. 2000).
The synthetic spectra are also indistinguishable in these two cases.
Unfortunately, many elements from the \zabs = 1.674 system
are represented by their upper limits.
Besides, we do not observe different ionic transitions
of the same element (like \ion{C}{iii} and \ion{C}{iv}, or \ion{N}{iii}
and \ion{N}{v}) which could help us to select the most adequate
ionizing continuum. However, the relative abundance ratio [C/O] = 
$-0.55\pm0.15$ 
measured for the model with the power law spectrum
shows formally
a better concordance with the [C/O] values observed in
the extragalactic metal-poor \ion{H}{ii} regions ([C/O] $\simeq -0.5$)
than the ratio [C/H] = 
$-0.72\pm0.15$
measured with the HM spectrum.
But this uncertainty in the shape of the ionizing spectrum does not
hamper the classification of the present system as a metal-poor
highly ionized absorber 
with the linear size of several kiloparsecs.

The question is whether this system is physically associated  
with the quasar environment or not. The virial mass of the absorber is
estimated as $M_{\rm vir} \simeq 10^9 M_\odot$ and 
if independent from the QSO, its proper luminosity 
distance is $D = 12.2 h^{-1}_{75}$ Mpc. Such masses are
comparable to that of dwarf galaxies whereas the 
distance of $\ga 10$ Mpc is 
consistent with distances between galaxies.
The measured metallicity [C/H] $\sim 1/40-1/130 Z_\odot$ 
is within the range
of metal abundances found in galactic halos. 
Thus, we may conclude that the most probable origin of 
the \zabs = 1.674 absorption system
is an external halo of some intervening galaxy 
located not far from the QSO and directly affected
by its radiation.

\subsection{\ion{O}{vi} absorber at \zabs = 1.602} 

There are several unsaturated hydrogen lines and
a weak \ion{O}{vi} 1031, 1037 \AA\, doublet in this system (see Fig.~7).
We also used 
continuum windows at the expected positions of 
\ion{C}{iii} 977 \AA, \ion{C}{iv} 1548 \AA, and \ion{N}{v} 1238 \AA\,
to constrain the ionization state and metallicity of this system,
and a window at the position of Ly$\gamma$ to
control the $N$(\ion{H}{i}) value.

The system was analyzed with the HM spectrum. The results obtained are
given in Col. (4), Table~1, the observed and overplotted synthetic
spectra are shown in Fig.~7. The volumetric gas density and the linear
size of the absorber were estimated for the value of $J_{912}$
given by HM for $z = 1.6$. Since the total number of ions is small,
the accuracy of the oxygen abundance
estimation is rather low, $\sim 50$\%.
The accuracy of the column density determinations are about 10\% and 20\%
for $N$(\ion{H}{i}) and $N$(\ion{O}{vi}), respectively.

\begin{table*}
\centering
\caption{Measured \ion{H}{i} column densities and upper limits (3 $\sigma$) 
for the metal-free systems toward HE 0515--4414
}
\label{tbl-1}
\begin{tabular}{llcccc}
\hline
\noalign{\smallskip}
\multicolumn{1}{c}{Parameter} & \multicolumn{1}{c}{$\lambda_0,$ \AA} &
\zabs = 1.38601 & \zabs = 1.49985 & 
\zabs = 1.66766 & \zabs = 1.68064  \\
\multicolumn{1}{c}{(1)} & \multicolumn{1}{c}{(2)} & (3) & (4) & (5) & (6) \\
\noalign{\smallskip}
\hline
\noalign{\smallskip}
$N$(H\,{\sc i}), \cm & Ly$\alpha,\beta,\gamma,\delta,\zeta$ & 1.1E15  &1.7E15  &2.7E14  &1.8E15  \\ 
$N$(C\,{\sc ii}), \cm  &1334.5 & $<$1.7E12 &$<$1.1E12  &$<$6.0E11  &$<$5.6E11  \\  
$N$(Si\,{\sc ii}), \cm  &1260.4, 1526.7& $<$6.7E11 &$<$9.6E11 &$<$7.1E10  &$<$9.0E10  \\ 
$N$(Fe\,{\sc ii}), \cm  &2382.8& $<$5.0E10 &$<$7.6E10 &$<$5.2E10  &$<$6.7E10  \\
$N$(C\,{\sc iii}), \cm  &977.0&$<$4.6E12 &$<$1.8E12 &$\ldots$   &$\ldots$  \\
$N$(Si\,{\sc iii}), \cm &1206.5&$<$5.9E11 &$<$8.0E11 &$<$1.2E11  &$\ldots$  \\
$N$(C\,{\sc iv}), \cm  &1548.2, 1550.8&$<$2.9E11  &$<$3.2E11 &$<$3.0E11  &$<$4.5E11  \\
$N$(Si\,{\sc iv}), \cm &1393.8, 1402.8&$<$2.3E11  &$<$2.2E11 &$<$9.7E10  &$<$7.8E10  \\ 
$N$(N\,{\sc v}), \cm  &1238.8, 1242.8&$<$7.1E12  &$<$1.8E12  &$<$2.2E12  &$<$1.0E12  \\ 
$N$(O\,{\sc vi}), \cm &1031.9, 1037.6&$<$1.8E13  &$<$4.0E12  &$<$7.5E12  &$<$3.3E13  \\ 
\noalign{\smallskip}
\hline
\end{tabular}
\end{table*}

Here we observe again
a low value of the Doppler parameter $b_{\rm O VI} = 12.5\pm3.6$
\kms\, (RBHL) 
which may indicate that the kinetic temperature is not
very high. The system exhibits a complex structure -- two strong subcomponents
are seen in the hydrogen Ly$\alpha$ profile. The widths of these components are
not consistent with temperatures larger than 10$^5$K. But the \ion{O}{vi} profiles might be
caused, in principle, by collisional ionization if one assumes that
another wide and shallow Ly$\alpha$ subcomponent is hidden in the spectrum.
Since, however, there is no evidence for a broad hydrogen component, 
it is not possible to carry out any quantitative estimations for this case.

According to the recovered values of the physical parameters
(high $U_0 = 0.26$, low $Z \la 1/10 Z_\odot$, low gas density
$n_0 \simeq 1.6\times10^{-4}$ \cmm, and the line of sight 
size $L \simeq 17$ kpc),
the \zabs = 1.602 system may be formed in the external halo of
an intervening galaxy.
The luminosity distance of $\simeq 38 h^{-1}_{75}$ Mpc 
implies that
this \ion{O}{vi} system is not influenced directly by the QSO
radiation.

\subsection{\ion{O}{vi} absorber at \zabs = 1.385} 

This system consists of at least two subcomponents
with very different chemical compositions:
a metal-rich component at \zabs = 1.3849 which reveals
absorption lines of 
\ion{C}{iii} 977 \AA,
\ion{C}{iv} 1548, 1550 \AA\, and
\ion{O}{vi} 1031 \AA\, (\ion{O}{vi} 1037 \AA\, is blended),
and a component shifted at $\Delta v \simeq 140$ \kms\,
and showing no 
clear metal absorption in the observational range
(see Fig.~8). 
The absorption features seen in the \ion{C}{iii} panel in Fig.~8
at $\Delta v \la -30$ \kms\, and $\Delta v \ga 40$ \kms\, belong to the
\ion{N}{ii} $\lambda1084$ \AA\, subcomponents from the damped Ly$\alpha$ system
at \zabs = 1.15 (de la Varga et al. 2000). 
Surprisingly the second component at $\Delta v \simeq 140$ \kms\,
is much more abundant in neutral hydrogen.

The metal-rich subsystem shows an unsaturated Ly$\alpha$ and no
distinguished absorption in Ly$\beta$. The measured 
neutral hydrogen column density is $N$(\ion{H}{i})  
$\simeq 5.3\times10^{13}$ \cm\, (Table~1, Col.~[3]). The
saturated Ly$\alpha$ and Ly$\beta$ lines of the metal-free subsystem 
give
$N$(\ion{H}{i}) $\simeq 1.1\times10^{15}$ \cm\, 
and $b = 35.4$ \kms,
that are estimated
from the Voigt profile fitting shown by the dotted lines in Fig.~8.

The \ion{O}{vi} system at \zabs = 1.385 was firstly
analyzed with the HM spectrum at $z = 1.4$. 
The obtained physical parameters are listed in Table~1, Col. (3).
In Fig.~8, the synthetic profiles (smooth lines) are shown
together with the 
observational data (dots with 1 $\sigma$ error bars) in a wide
radial velocity range to illustrate different metal composition in the
metal-rich and metal-free subsystems. 

We note that the accuracy of the fit to the blue wing
of the Ly$\alpha$ is not sufficiently high.
The calculated profile lies slightly over the data points in the range
$-55$ \kms\, $\la \Delta v \la -30$ \kms\, and under the observed
intensities in the range
$-15$ \kms\, $\la \Delta v \la 0$ \kms. However, 
the shape of the synthetic Ly$\alpha$ profile is fixed by the 
spectral shapes of the \ion{C}{iv} lines which were observed with
considerably higher S/N  and higher spectral resolution. Any 
attempts to improve the fitting of the blue wing of the Ly$\alpha$
led to distortions in the synthetic \ion{C}{iv} profiles inconsistent
with the data points. For the same reason the red wing of the Ly$\alpha$
cannot be self-consistently fitted to the model in the range
$15$ \kms\, $\la \Delta v \la 70$ \kms\, 
under the assumption of a constant metallicity
across the absorber. 
If, however, we do observe a strong metallicity gradient,
then the fitting of the Ly$\alpha$
may be easily improved assuming an additional metal-free component 
located at $\Delta v \simeq 30$ \kms. Higher S/N data are required to
investigate this spectrum in more detail.
Within the present model, we may
overestimate the value of $N$(\ion{H}{i}) at
\zabs = 1.3849 by a few percent.

The measured values of $U_0$, metallicity, $n_0$, and
$T_{\rm kin}$ are very much alike to those found in the \zabs = 2.848
system toward Q0347--3819 (Levshakov et al. 2003a).
We consider this similarity as a support 
for our assumption that these systems may be high-$z$ counterparts
of the HVCs observed in the Milky Way halo. 

\begin{figure*}
\vspace{0.0cm}
\hspace{0.0cm}\psfig{figure=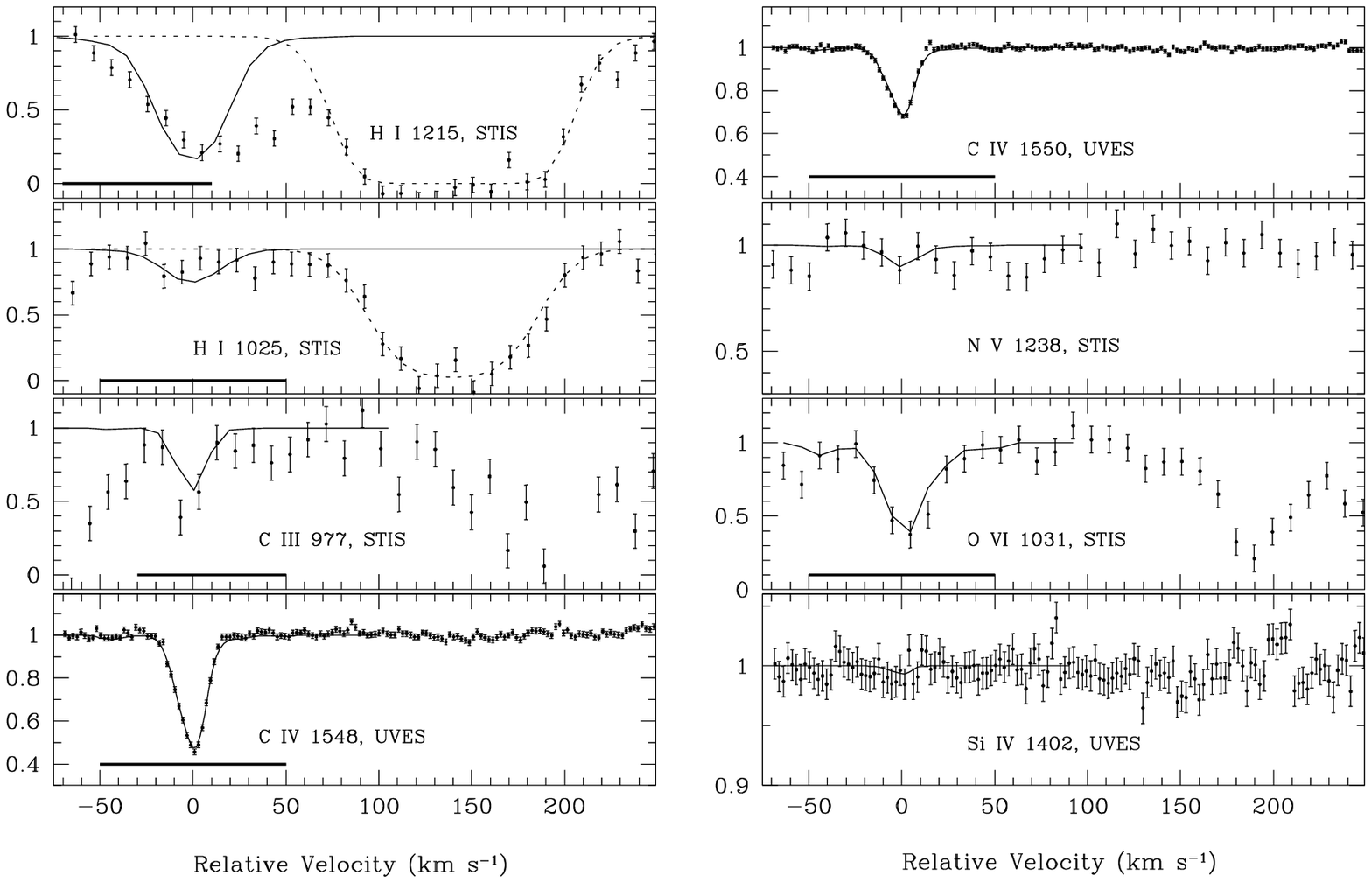,height=15.0cm,width=18.0cm}
\vspace{-6.5cm}
\caption[]{Same as Fig.~4 but for the \zabs = 1.385 \ion{O}{vi} 
absorption system.
The zero radial velocity is fixed at $z = 1.3849$. 
The corresponding physical parameters are listed in Table~1, 
Col. (3). Here
$\chi^2_{\rm min} = 1.20$, $\nu = 123$.
In addition to the \zabs = 1.385 system, hydrogen absorption seen at
$\Delta v \simeq 140$ \kms\, is shown by dotted lines. No metals are
detected at this $\Delta v$. 
A wide absorption feature 
in the \ion{O}{vi} panel at $\Delta v > 130$ \kms\, is due to
the \ion{Fe}{ii} $\lambda1144$ \AA\, line from the damped Ly$\alpha$ system
at \zabs = 1.15 (de la Varga et al. 2000), whereas absorption features
in the \ion{C}{iii} panel at $\Delta v \la -30$ \kms\, and
$\Delta v \ga 40$ \kms\, are caused by the \ion{N}{ii} $\lambda1084$ \AA\,
absorption components from the \zabs = 1.15 system. 
}
\label{fig8}
\end{figure*}

A relatively low ratio of [C/O] = $-0.38\pm0.17$ at the measured oxygen
abundance of 12+log(O/H) = 8.65 may, however, indicate that
the choice of the HM background was not optimal. We tried to investigate this
question by using slightly modified HM spectra. 
As noted by Fardal et al. (1998),
in calculated metagalactic UV spectra 
the strength of the break at 4 Ryd 
(which accounts for \ion{He}{ii} attenuation in intervening clouds)
depends strongly on the assumed opacity model at a given redshift.
We found that the derived [C/O] ratio 
is very sensitive to
small changes in the value of this decrement at 4 Ryd. 
For instance, if we modify the HM continuum by
decreasing the depth of the break at 4 Ryd by a 
factor of 1.25 (0.1 dex), 
then [C/O] = $-0.12\pm0.17$ which is in a much better
concordance with the measurements in the extragalactic \ion{H}{ii}
regions. The corresponding physical parameters,
calculated with this modified HM spectrum, are listed in Table~1,
Col. (2) (the accuracy of $Z_{\rm C}$ and $Z_{\rm O}$ is
about 15\% and 40\%, respectively).

The choice of the photoionization model for the \zabs = 1.385 system
can be supported by the same arguments as for the \ion{O}{vi}
cloud at \zabs = 1.674. Namely, we observe
$b_{\rm C IV}/b_{\rm O VI} \simeq 0.6$ (RBHL) as well as 
a relatively large ratio
$N$(\ion{C}{iv})/$N$(\ion{O}{vi}) $\simeq 0.2$ which 
cannot be realized in collisionally ionized gas.

We can also set an upper limit on the metallicity in the metal-free cloud 
using the  estimated limits from Table~2.
Because of the blended range at $\Delta v > 50$ \kms, we are not able to judge
whether a \ion{C}{iii} $\lambda977$ \AA\, absorption is present at
$\Delta v \simeq 150$ \kms\, or not. If we attribute the absorption feature at
$\Delta v \simeq 150$ \kms\, to the \ion{C}{iii} line, then it would require
a high $\Upsilon_{\rm CIII}/\Upsilon_{\rm CIV}$ ratio and, hence, a low ionization
parameter, $U_0 \sim 10^{-3}-10^{-2}$, which gives the gas number density of
$n_0 \sim 10^{-2}-10^{-3}$ \cmm\, and a cloud size
from 10s to 100s of pc, assuming the HM background spectrum. 
The corresponding carbon abundance limit would be [C/H] $<-1.5$ and $<-2.5$.
However, the \ion{O}{vi} cloud at $\Delta v = 0$ \kms\, has the linear size of
$\sim 2$ kpc which is too low for a cloud to be in space on its own, so both
systems are probably physically connected. 
Taking into account that the \ion{O}{vi} system exhibits a high
metallicity, the structure of two clouds when a metal-poor component
($\Delta v = 150$ \kms) is surrounded and 
pressure confined by a hot metal-rich gas
($\Delta v = 0$ \kms) seems to be quite unphysical.
A more plausible interpretation of the observed spectrum is that the line of sight
intersects some peculiar object (like a star burst region, a superbuble, an infalling shell
fragment etc.) which is metal-enriched and embedded in a metal-deficit halo of an
intervening galaxy. This metal-poor gas  
is likely to be photoionized since the estimated Doppler
parameter $b_{\rm H} = 35.4$ \kms\, implies $T \leq 7.4\times10^4$ K.
If we assume that 
the incident UV radiation has the HM spectrum at $z = 1.4$ and
the ionization state in the metal-poor system is
approximately the same as in the nearest metal-rich system
(i.e., $U_0 \ga 0.1$), then for a constant gas density
($\sigma_{\rm y} = 0$) the conservative 
limits on the total oxygen and carbon column densities are 
$N$(O) $< 1.8\times10^{14} $ \cm\, and
$N$(C) $< 2.0\times10^{12}$ \cm, while the total hydrogen column density is 
$N$(H) $= 4.4\times10^{19}$ \cm. 
Under this assumption, we obtain [O/H] $< -{2.0}$ and [C/H] $< -3.7$ and
the linear size of this absorber $L \ga 90$ kpc. The upper limits on
the metal abundances may be increased by about 0.3--0.5 dex
and the linear size decreased by several times, 
if the density is inhomogeneous (e.g., $\sigma_{\rm y} \sim 1$). 

The metallicity of the pristine
gas at the level $10^{-4}$ to $10^{-3}Z_\odot$ can be 
already produced by the first 
Population III stars (e.g., Nakamura \& Umemura 2001). At face value,
our estimations suggest that this metal-free 
system can indeed consists of  primordial gas. 

\begin{figure*}
\vspace{0.0cm}
\hspace{0.0cm}\psfig{figure=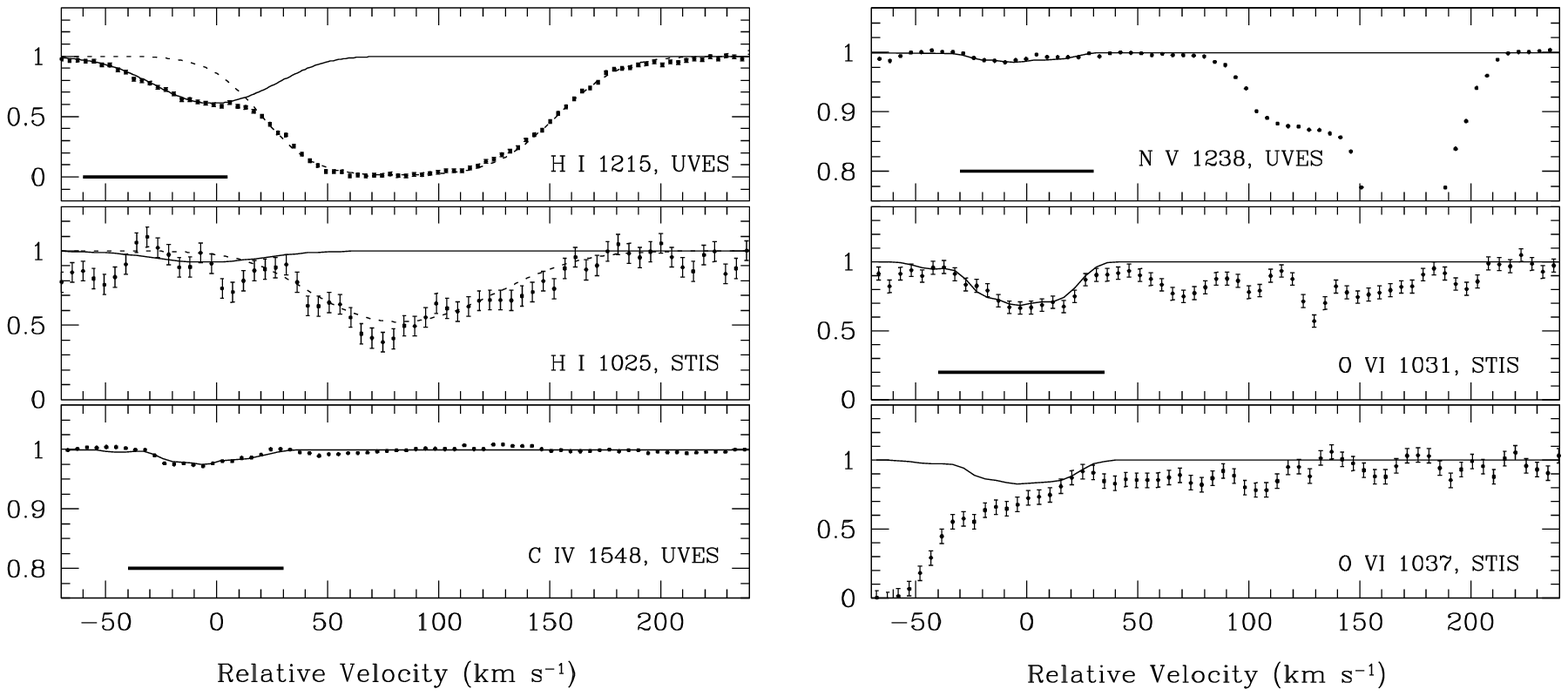,height=15.0cm,width=18.0cm}
\vspace{-9.0cm}
\caption[]{Same as Fig.~4 but for the \zabs = 1.667 \ion{O}{vi} 
absorption system.
The zero radial velocity is fixed at $z = 1.66695$. 
The corresponding physical parameters are listed in Table~1, 
Col. (5). Here
$\chi^2_{\rm min} = 0.91$, $\nu = 99$.
In addition to the \zabs = 1.667 system, hydrogen absorption seen at
$\Delta v \simeq 80$ \kms\, is shown by dotted lines. No metals are
detected at this $\Delta v$. The absorption features in the range 
$\Delta v \simeq 125-200$ \kms\, in the \ion{O}{vi} 1031 \AA\, panel are caused by 
the \ion{C}{i} 1280 \AA\, lines 
from the damped Ly$\alpha$ system at \zabs = 1.15 (de la Varga et al. 2000),
whereas features at $\Delta v \simeq 70$ \kms\, and 100 \kms are unidentified.
}
\label{fig9}
\end{figure*}

\subsection{\ion{O}{vi} absorber at \zabs = 1.667} 

This new \ion{O}{vi} system identified in the present study
reveals the striking similarity to the preceding \ion{O}{vi}
absorber at \zabs = 1.385.
Two close subcomponents with very different chemical compositions
are observed~: a metal-rich component at \zabs = 1.66695 showing
absorption lines of \ion{C}{iv} 1548 \AA, \ion{N}{v} 1238 \AA\, and
\ion{O}{vi} 1031 \AA\, (\ion{C}{iv} 1550 \AA\, and \ion{N}{v} 1242 \AA\,
are too weak, and \ion{O}{vi} 1037 \AA\, is partly blended as seeing in Fig.~9),
and a component shifted at $\Delta v \simeq 80$ \kms\, without any
distinctive
metal absorption 
 both in low (\ion{C}{ii}, \ion{SiII}, \ion{Fe}{ii}) and
high (\ion{Si}{iii}, \ion{Si}{iv}, \ion{C}{iv}, \ion{N}{v}, \ion{O}{vi})
ionic transitions. The \ion{C}{iii} line is blended with strong Ly$\gamma$ absorption
from the metal-free system at \zabs = 1.681.
The second component shows again considerably higher
neutral hydrogen column density, $N$(\ion{H}{i}) $\simeq 2.7\times10^{14}$ \cm\,
(Table~2, Col. [5])
as compared to the metal-rich system with 
$N$(\ion{H}{i}) $\simeq 2.2\times10^{13}$ \cm (Table~1, Col. [5]).

The observed Ly$\alpha$ and Ly$\beta$ profiles are shown 
in Fig.~9 where the overplotted solid lines are the MCI
solutions and the dotted lines correspond to the Voigt profiles
(centered at $\Delta v = 67.6$ \kms\, and 100.0 \kms) which were
used to estimate the total $N$(\ion{H}{i}) in the metal-free component.

The metal line profiles are rather wide and shallow (shown by dots in Fig.~9 
are their profiles filtered with a median filter to increase the
contrast) and exhibit non-Gaussian shapes.  
Their apparent widths are
FWHM$_{\rm C IV} \simeq$ FWHM$_{\rm O VI} \simeq 45\pm3$ \kms\,
(\ion{N}{v} is too weak to provide an accurate estimation), and that
of the hydrogen component at $\Delta v = 0$ \kms\,
is FWHM$_{\rm H I} \simeq 56\pm3$ \kms.
Although the widths of \ion{C}{iv} and \ion{O}{vi} allow formally for
the temperature $T_{\rm kin} > 10^5$ K,
their collisional ionization should nevertheless be ruled out
because the hydrogen Ly$\alpha$ line is not wide enough  
and its profile cannot be extended over $| \Delta v | > 50$ \kms\, due to
high S/N in both hydrogen ( $\Delta v < -50$ \kms) and carbon 
($\Delta v > 50$ \kms) lines.
On the contrary, the observed profiles of these hydrogen and metal
lines can be well described assuming turbulent broadening and photoionization by
a single power law continuum $J_\nu \propto \nu^{-1.5}$
normalized at 1 Ryd to $J_{\rm Q} = 2\times10^{-21}$
erg cm$^{-2}$ s$^{-1}$ Hz$^{-1}$ sr$^{-1}$. 
The corresponding 
physical parameters are given in Table~1, Col. (5).
The accuracy of the estimated metallicities is not high ($\simeq \pm 0.2$ dex) 
since carbon and nitrogen
lines are very weak and the signal-to-noise ratio in the \ion{O}{vi}
data is rather low. But the most probable values of
[C/O] $\simeq -0.1$ and [N/O] $\simeq -0.4$ are in line with the
abundances measured in the extragalactic \ion{H}{ii} regions
with a relatively high metallicity 12 + log(O/H) $\simeq 8.2$
(cf., Henry et al. 2000; Centuri\'on et al. 2003).

\begin{figure*}
\vspace{0.0cm}
\hspace{0.0cm}\psfig{figure=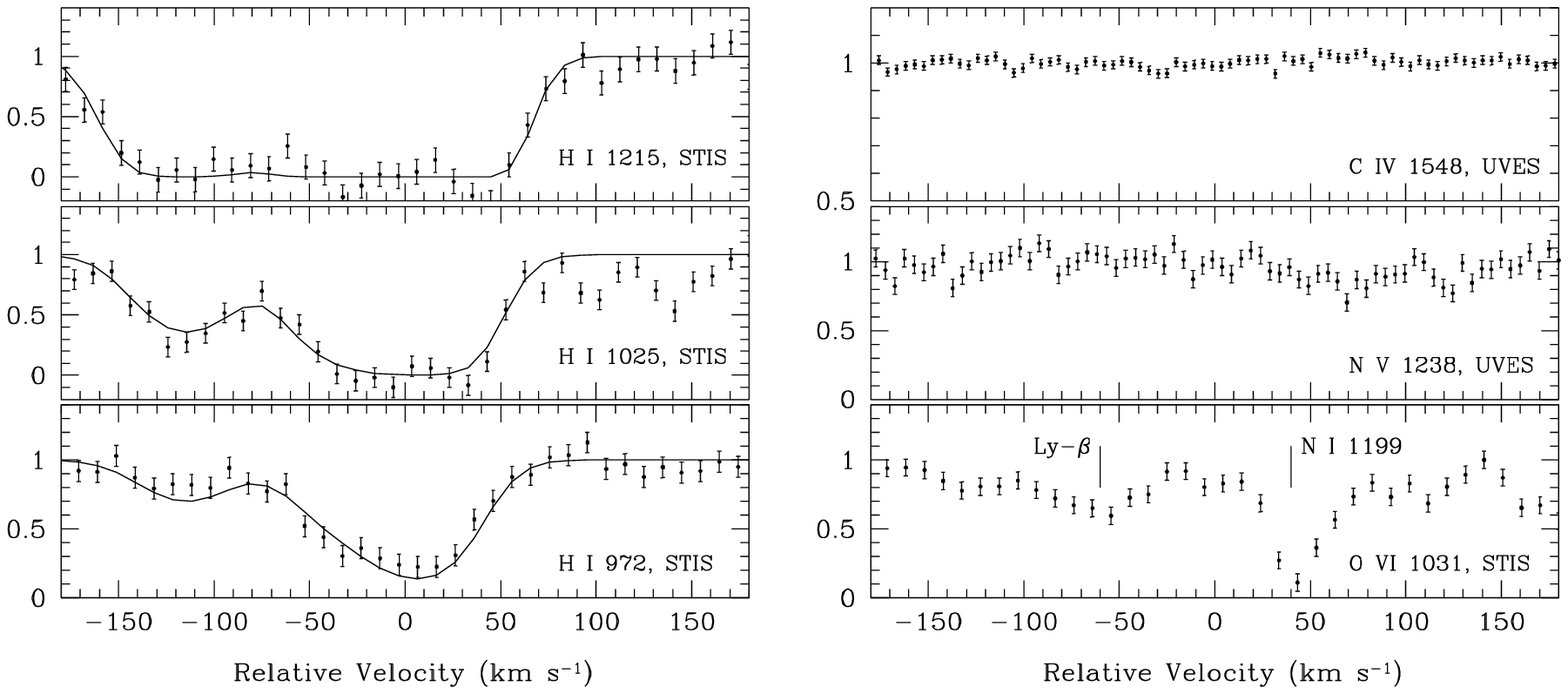,height=14.0cm,width=18.0cm}
\vspace{-8.5cm}
\caption[]{An example of the metal-free Ly$\alpha$ system at
\zabs = 1.500 toward HE 0515--414. 
The zero radial velocity is fixed at $z = 1.499854$. 
Smooth lines are the combined synthetic spectra of three
subcomponents centered at
$\Delta v = -114.5$, $-27.6$, and 11.0 \kms\, and convolved
with the STIS point-spread function. The cental components show the
total $N$(\ion{H}{i})  $\simeq 1.8\times10^{15}$ \cm.
In panel \ion{O}{vi},
the absorption features at $\Delta v \simeq
-60$ \kms and $\simeq 40$ \kms are due to the
Ly$\beta$ line from the \zabs = 1.515 system and to 
the \ion{N}{i} 1199 \AA\, line
from the damped Ly$\alpha$ system
at \zabs = 1.15 (de la Varga et al. 2000).
}
\label{fig10}
\end{figure*}

The obtained metallicity $Z \simeq 1/3 Z_\odot$,
number density $n_0 \simeq 4\times10^{-4}$ \cmm\,
and the linear size $L \simeq 900$ pc are typical for the HVCs
observed in the Milky Way. We suggest that this 
system may be embedded in an external halo of some galaxy at
\zabs $\simeq 1.667$. 
The wide and shallow metal lines may be explained 
if the metal-enriched cloud is driven away by an outflow similar to that
observed in the wings of the Ly$\alpha$ line from the neighbour system at
\zabs = 1.674. 
Using the mean ionization
parameter $U_0 \ga 0.2$ for the metal-free system 
at $\Delta v \simeq 80$ \kms\,
and assuming that it
is photoionized by the same background spectrum
(i.e. $J_\nu \propto \nu^{-1.5}$), 
we obtain a
conservative upper limit for the metal abundances in this galactic
halo: [C/H] $< -2.4$ and [O/H] $< -2.5$ (upper limits on the column
densities from Table~2, Col. [5] are used).

Thus, both \ion{O}{vi} systems, that at \zabs = 1.385 and 
\zabs = 1.667, show a strong metallicity gradient
$|$d[C/H]/d$v|$ $> 0.02$ (\kms)$^{-1}$, which may imply
that the metal enrichment of the metagalactic
medium was very inhomogeneous in space and time. 
Ly$\alpha$ systems described
in the next section support this conclusion.

\subsection{Metal-free systems at \zabs = 1.500 and 1.681}

The metal-free absorbers showing a moderate neutral hydrogen column
density of $N$(\ion{H}{i}) $\sim 10^{15}$ \cm\, are not rare in QSO
spectra. For instance, in HE 0515--4414 we identified another two 
systems at \zabs = 1.500 and 1.681
with, respectively, $N$(\ion{H}{i}) $\simeq 1.7\times10^{15}$ 
and $1.8\times10^{15}$ \cm\, 
(the uncertainty is about 6\% for both cases)
without any clear metal lines in the observational wavelength range.

Fig.~10 represents the \ion{H}{i} Ly$\alpha$,
Ly$\beta$, and Ly$\gamma$ lines,
and the continuum windows at the expected positions
of the \ion{C}{iv} 1548 \AA, \ion{N}{v} 1238 \AA, and \ion{O}{vi} 1031 \AA\,
lines from the former system.
We find that three components centered at
$\Delta v = - 114.5$, $-27.6$, and 11.0 \kms\, describe adequately
the hydrogen absorption with $N$(\ion{H}{i}) = 
$2.7\times10^{14}$, $6.2\times10^{14}$, and $1.2\times10^{15}$ \cm,
and $b = 31$, 34, and 29 \kms, respectively (the combined synthetic
spectra are shown by the smooth lines). 

In Fig.~11, another metal-free system is shown.
Here three hydrogen components centered at
$\Delta v = -97.0$, $-34.0$, and 0.0 \kms\,
with $b = 54.7$, 48.0, and 22.0 \kms\, and
$N$(\ion{H}{i}) = $1.6\times10^{13}$, $5.25\times10^{14}$, and
$1.13\times10^{15}$ \cm, respectively, fit perfectly the
observed hydrogen profiles. However, we do not detect any
absorption in \ion{C}{iv} 1548 \AA, \ion{N}{v} 1238 \AA, and
\ion{O}{vi} 1031 \AA\, lines 
or in the low ions listed in Table~2, Cols. (4) and (6).
To deduce the upper limits on the metallicity for both systems, 
we applied the same assumptions as for
the metal-free 
systems previously described, i.e. photoionization by either the HM spectrum at
$z = 1.5$ or by the power law spectrum ($\nu = -1.5$)
at $z = 1.68$, the mean ionization parameter
$U_0 \la 0.1$ and the upper limits on the column densities from Table~2.
The estimated limits on the metal abundances are [C/H] $< -4.0$,
[O/H] $<-3.0$ (\zabs = 1.500) and [C/H] $< -3.0$, [O/H] $< -2.5$ (\zabs = 1.681),
assuming constant density.
 
We note that extremely low metal abundances are
detected not only in typical Ly$\alpha$ forest clouds with
$N$(\ion{H}{i}) $\sim 10^{15}$ \cm, but also in the so-called Lyman-limit
systems (LLSs) having $N$(\ion{H}{i}) $\ga 10^{17}$ \cm.
For instance, in the LLS at \zabs = 2.917 with
$N$(\ion{H}{i}) $= 3.2\times10^{17}$ \cm, the carbon abundance at the
level of $\simeq 0.001Z_\odot$ was recently measured (Levshakov et al. 2003b).

\begin{figure*}
\vspace{0.0cm}
\hspace{0.0cm}\psfig{figure=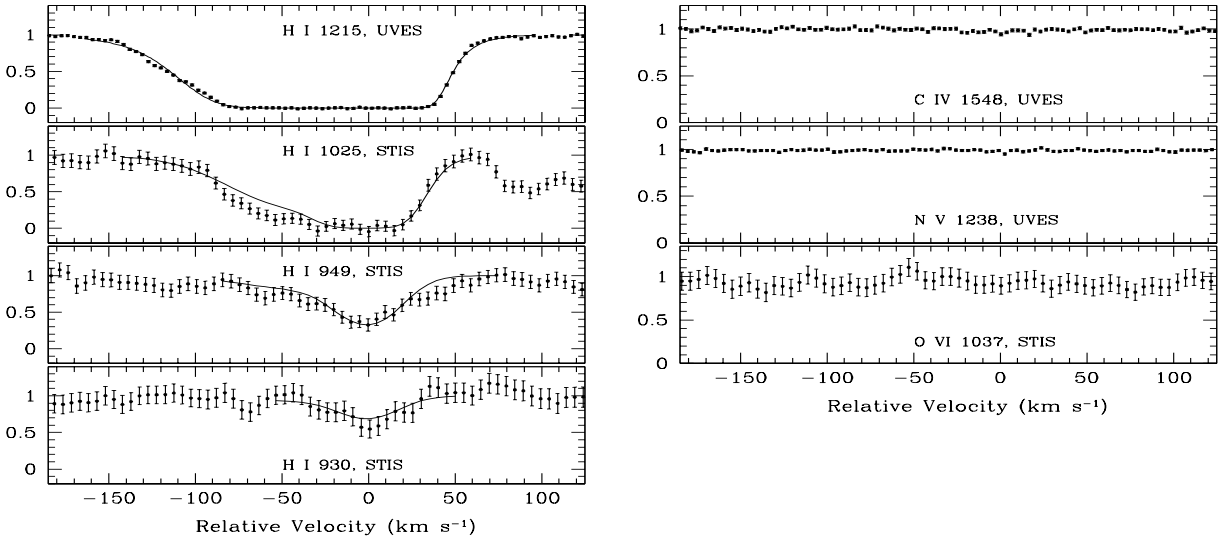,height=14.0cm,width=18.0cm}
\vspace{-7.0cm}
\caption[]{Same as Fig.~10 but for the metal-free Ly$\alpha$ system at
\zabs = 1.681. 
The zero radial velocity is fixed at $z = 1.68064$. 
Smooth lines are the combined synthetic spectra of three
subcomponents centered at
$\Delta v = -97.0$, $-34.0$, and 0.0 \kms\, and convolved
with the UVES and STIS point-spread functions. The total neutral hydrogen
column density is $N$(\ion{H}{i})  $\simeq 1.7\times10^{15}$ \cm.
In panel \ion{H}{i} 1025 \AA, the portion of the Ly$\beta$ line at
$\Delta v \simeq -60$ \kms is probably 
contaminated by the Ly$\alpha$ forest.
}
\label{fig11}
\end{figure*}

\section{Discussion}

We detected no prominent imprints of hot ($T_{\rm kin} > 10^5$ K), 
collisionally ionized gas in the analyzed \ion{O}{vi} systems.
All of them can be described self-consistently under the assumption of
photoionization only. 

Our analysis has shown that the \ion{O}{vi} absorbers toward the QSO
HE 0515--4414 originate in different environments:
the \zabs = 1.697 system is likely to be physically associated 
with the QSO/host galaxy,
two systems --- that at \zabs = 1.674 and 1.602 --- are 
formed in a smoothly distributed gas 
in the outer parts of the intervening galaxy halos, whereas the \zabs = 1.385
and \zabs = 1.667
absorbers are HVC-type clouds embedded in galaxy halos of an extremely low
metallicity ([C/H] $< -2.5$). 

The presented \ion{O}{vi} systems demonstrate 
a wide range of metal abundances: $-1.6 <$ [C/H] $< 0.6$. 
Moreover, we detected four systems with
$N$(\ion{H}{i}) $\sim 10^{15}$ \cm\, which do not show any metal
transitions in the observational wavelength range.

These facts imply that our sample of the \ion{O}{vi} absorbers 
does not trace the `warm-hot' gas 
indicating that its fraction is probably negligible 
between $z = 1.38$ and $z = 1.7$. 
Another output is that the present sample 
cannot be used for the estimation of the cosmological mass density of  
metals and baryons since the analyzed \ion{O}{vi} systems
are very heterogeneous and do not allow to make estimations of the
mean characteristics of the \ion{O}{vi}-bearing gas in general.
However, just this diversity of the considered \ion{O}{vi} systems
leads to the important conclusions concerning the mechanisms of
the metal enrichment of the IGM.

The most metal-rich system
at \zabs = 1.697 is an associated system
with metallicity of $\sim 5$ times solar. 
As mentioned in Sect.~4.1,
such high over solar abundances are usually 
observed in quasar broad emission
line regions. For example, the metallicity of $Z \sim 2Z_\odot$ was
measured in the \zem = 4.16 QSO BR 2248--1248 (Warner et al. 2002)
and the metallicities of $Z \sim 4Z_\odot$ were detected in the
spectra of 11 high redshift QSOs ($3.9 \la z \la 5.0$) by Dietrich et al. (2003).

According to galactic chemical evolution models, high metal abundances in the
gaseous environments near QSOs are accumulated over a period of
$\sim 1$ Gyr before the onset of quasar activity 
(e.g., Matteucci \& Padovani 1993;
Hamann \& Ferland 2000; Romano et al. 2002).
A galactic wind may be triggered at the point of the quasar's 
`turn on' due to the 
combined effect of QSO and stars. Calculations show that the more
massive galactic nuclei are, the shorter 
is time interval between the onset of the
star formation and the peak of the QSO activity (e.g., Monaco et al. 2000).

The mass of the central source of the quasar HE 0515--4414 can be
estimated from its luminosity
${\cal L}_{\rm Q} \simeq 2.5\times10^{47}$ erg s$^{-1}$ = 
$6.5\times10^{13}{\cal L}_\odot$.
Since it is the brightest QSO, we can assume that its luminosity
${\cal L}_{\rm Q}$ is close to the Eddington limit
${\cal L}_{\rm E} = 1.3\times10^{38} 
(M_{\rm Q}/M_\odot)$ erg s$^{-1}$, and thus
$M_{\rm Q} \simeq 2\times10^9 M_\odot$.

The fact that the ionization state of this metal-rich \ion{O}{vi} system
is maintained by a single power law spectrum $J_\nu \propto \nu^{-1.5}$
implies that the gas between the quasar and the absorbing cloud is highly
ionized since the incident UV continuum is not blocked by hydrogen and helium
absorption. 

A quasar with the luminosity of $6.5\times10^{13} {\cal L}_\odot$ and the
ionizing continuum $\propto \nu^{-1.5}$ can essentially ionize gas in a 
radius of about 1.1 Mpc, assuming $n_{\rm H} = 0.01$ \cmm\, and the
relative helium abundance (by number) of 10\%.
So, the distance between the \ion{O}{vi} cloud and the QSO nucleus is to be
$R_{\rm cl} \la 1$ Mpc. If the gas in this cloud  
has been enriched by heavy elements
synthesised by earlier ($\Delta t \sim 1$ Gyr) star formation episode
after which
the cloud was blown out into the outer zone of the galactic halo, then the
mean velocity of the outflow 
should be $R_{\rm cl}/\Delta t \sim 1000$ \kms, which is
comparable to the observed velocity difference $\Delta v_{\rm em-abs}$.
An outflow speed as high as 1000 \kms\, has been observed in the Lyman
break galaxy MS 1512--cB58 at $z = 2.7276$ which is undergoing active
star formation (Pettini et al. 2002).

If the \ion{O}{vi} system is located at 1 Mpc from the QSO, then its linear
size $L$ and the mean gas density $n_0$ 
should be scaled by a factor of $4.5^2$
as compared to the values from Table 1, Col. (7),  
giving $L \simeq 0.8$ pc
and $n_0 \simeq 0.2$ \cmm. Thus, the \zabs = 1.697 system may 
very well be an
example of a thin shell or a fragment of a superbubble whose
expansion is probably caused by a cumulative effect of supernova
and the quasar activity. Such metal-rich outflowing gas
is supposed to 
give rise to some HVCs observed in the
Milky Way halo (Bregman 1980; Wakker 2001).
  
Other metal-rich \ion{O}{vi} systems 
at \zabs = 1.385 ($Z \sim 1/3 Z_\odot$, $L \simeq 2$ kpc)
and at \zabs = 1.667 ($Z \sim 1/3 Z_\odot$, $L \simeq 0.9$ kpc)
may be produced by 
a mechanism known as a galactic fountain
(Bregman 1980):
gas contaminated by heavy elements arises from the inner region 
of an intervening  galaxy and condenses into a cloud within the halo.
After formation, the cloud cools and falls back toward the galaxy center. 
This mechanism was suggested to explain the origin of the high-metallicity
HVCs in the Milky Way. 

We can assume that the galactic fountain also
functions in distant galaxies.
In our case we observe extremely metal-poor halos
([C/H] $< -3$) 
which means that
star formation processes were not very powerful and therefore
the mass of the host galaxy was not, probably, very high.
Metal-free environments can also indicate that the HVC is situated in an
external part of the halo, which again could be possible only if the
gravitational potential of the galaxy was rather low.
However, we cannot exclude the possibility that  the origin of 
these \ion{O}{vi} clouds is extragalactic.
For some Galactic HVCs there are indications that they
are located at
typical distances from the Milky Way of up to 1 Mpc and currently falling
into the Local Group (e.g., Blitz et al. 1999).
It was shown 
above that a superbubble fragment like the \zabs = 1.697
cloud can be blown away from the host galaxy 
to the distance of $~\sim 1$ Mpc. 
This distance is large enough for a fragment to become a `loose' cosmic
object which eventually can be captured by an encountered galaxy.

In the redshift interval from $z = 1.385$ to $z = 1.697$ we have found
four metal-free systems at \zabs = 1.385, 1.500, 1.667, and 1.681 which
lie among five metal-rich systems at \zabs = 1.385, 1.602, 1.667, 1.674, and
1.697. All metal-free systems reveal rather high neutral
hydrogen column densities~: $1.8\times10^{15}$ \cm\, (\zabs = 1.500),
$1.7\times10^{15}$ \cm\, (\zabs = 1.681),
$1.1\times10^{15}$ \cm\, (\zabs = 1.385), and
$2.7\times10^{14}$ \cm\, (\zabs = 1.667). 
Taking this into account we may conclude that our results
favour {\it in situ} enrichment scenarios which were proposed in a number
of recent publications (see, e.g., Scannapieco et al. 2002 and references
therein).

\section{Summary} 

We have deduced the physical properties of the five \ion{O}{vi} systems
and four Ly$\alpha$ metal-free systems
in the range $\Delta z = 1.385 - 1.697$ toward HE 0515--4414. The main
conclusions are as follows:
\begin{enumerate}
\item All \ion{O}{vi} systems can be self-consistently 
described under the assumption of 
photoionization equilibrium only. 
This implies that the fraction of shock-heated
hot gas with temperature $T_{\rm kin} > 10^5$ K is negligible in these systems.
\item The analyzed \ion{O}{vi} systems 
belong to a {\it heterogeneous} population
which is formed by at least three groups of absorbers~: 
($i$) gas in a thin shell of a superbubble associated with the QSO/host
galaxy (\ion{O}{vi} at \zabs = 1.697);
($ii$) extended low metallicity gas
halos of distant galaxies (\ion{O}{vi} at \zabs = 1.674 and 1.602); and
($iii$) metal-enriched gas arising from the inner
galactic regions or falling into the external galactic halo 
(\ion{O}{vi} at \zabs = 1.385 and 1.667).
\item Only a power law type spectrum of the ionizing UV radiation is
consistent with the observed sample of metal lines from the associated
\zabs = 1.697 system. The optimal fitting was found with the spectral
index $\alpha = -1.5$. The measured metal abundances are about 5 times solar,
and the metallicity pattern is solar. The system, 
located at $\sim 1$ Mpc from the QSO, completely covers the
continuum source. 
The line-of-sight size of the system is $L \simeq 0.8-16$ pc. 
\item The absorption systems at \zabs = 1.385 and 1.667 show  
characteristics very
similar to that observed in metal-enriched HVCs in the Milky Way. These
systems can be interpreted as high-redshift counterparts of 
Galactic HVCs.
\item An important outcome of our study 
is that Ly$\alpha$ absorbers are utterly inhomogeneous in
metal abundances. 
Upper limits on metal contents at the
extremely low level of $Z < 10^{-3}Z_\odot$ were set for
three systems with $N$(\ion{H}{i}) $\sim 10^{15}$ \cm.
\end{enumerate}

\begin{acknowledgements}
S.A.L. gratefully acknowledges the hospitality of
Hamburger Sternwarte, Universit\"at Hamburg.
We thank our referee for helpful comments.
The analysis of the HST data has
been supported by the Verbundforschung of the BMBF/DLR under grant
No. 50 OR 0203.
The work of S.A.L. and I.I.A. is supported in part by the
RFBR grant No.~03-02-17522.
\end{acknowledgements}

\end{document}